\documentclass[fleqn,usenatbib]{mnras}

\usepackage[T1]{fontenc}
\usepackage[table,xcdraw]{xcolor}

\usepackage{booktabs}

\DeclareRobustCommand{\VAN}[3]{#2}
\let\VANthebibliography\thebibliography
\def\thebibliography{\DeclareRobustCommand{\VAN}[3]{##3}\VANthebibliography}

\usepackage{graphicx}	
\usepackage{amsmath}	
\usepackage{amssymb}	
\usepackage{subcaption}

\usepackage{hhline}

\usepackage{newtxtext,newtxmath}

\title[MeerTRAP: Galactic fast transients]{MeerTRAP: Twelve Galactic fast transients detected in a real-time, commensal MeerKAT survey}

\author[M. C. Bezuidenhout et al.]{M. C. Bezuidenhout,$^{1}$\thanks{E-mail: mechiel.bezuidenhout@postgrad.manchester.ac.uk}
E.~Barr$^2$,
M.~Caleb$^{1}$,
L.~N.~Driessen$^{3}$,
F.~Jankowski$^{1}$,
M.~Kramer$^2$,
M.~Malenta$^{1}$,\newauthor
V.~Morello$^{1}$,
K.~Rajwade$^{1}$,
S.~Sanidas$^{1}$,
B.~W.~Stappers$^{1}$,
and M.~Surnis$^{1}$
\vspace{0.5cm}
\\
$^{1}$Jodrell Bank Centre for Astrophysics, University of Manchester, Manchester M13 9PL, UK\\
$^{2}$Max-Planck-Institut f\"ur Radioastronomie, Auf dem Hugel 69, D-53121 Bonn, Germany\\
$^{3}$CSIRO, Space and Astronomy, PO Box 1130, Bentley, WA 6102, Australia
}

\date{Accepted XXX. Received YYY; in original form ZZZ}

\pubyear{2022}

\begin{document}
\label{firstpage}
\pagerange{\pageref{firstpage}--\pageref{lastpage}}
\maketitle

\begin{abstract}
MeerTRAP is a real-time untargeted search project using the MeerKAT telescope to find single pulses from fast radio transients and pulsars. It is performed commensally with the MeerKAT large survey projects (LSPs), using data from up to 64 of MeerKAT's 13.96~m dishes to form hundreds of coherent beams on sky, each of which is processed in real time to search for millisecond-duration pulses. We present the first twelve Galactic sources discovered by MeerTRAP, with DMs in the range of 33--381~pc~cm$^{-3}$. One source may be Galactic or extragalactic depending on the Galactic electron density model assumed. Follow-up observations performed with the MeerKAT, Lovell, and Parkes radio telescopes have detected repeat pulses from seven of the twelve sources. Pulse periods have been determined for four sources. Another four sources could be localised to the arcsecond-level using a novel implementation of the tied-array beam localisation method.
\end{abstract}




\section{Introduction}

Transient astrophysical radio sources are typically considered to be those that change appreciably on short time scales. Involving vast amounts of energy released in a short span of time, these are naturally associated with some of the most extreme environments in the universe in terms of quantities such as density, pressure, temperature, velocity, and gravitational and magnetic field strengths. \textit{Fast} radio transients are even more extreme, varying on time scales of a second or less \citep[][]{Cordes2007}. Until recently, high time resolution single pulse searches have been confined to single-dish radio telescopes with large FoVs to ensure high survey speeds. This comes at the expense of the ability to precisely localise sources. However, in the coming era of large-scale, interferometry-based observatories such as the Square Kilometre Array \citep[SKA;][]{2009dewdney}, large FoV fast radio transient surveys with excellent potential for localisation are becoming a reality. Ongoing surveys with SKA precursors like MeerKAT \citep[][]{2016jonas} in South Africa, the Low Frequency Array \citep[LOFAR;][]{2013vanhaarlem} in the Netherlands, and the Murchison Widefield Array \citep[MWA;][]{2013tingay} and Australian Square Kilometre Array Precursor \citep[ASKAP;][]{2008Johnston} in Australia are already demonstrating the utility of such telescopes for these purposes. Fast transient searches with other interferometric telescopes like the Canadian Hydrogen Intensity Mapping Experiment \citep[CHIME;][]{2018chime} and the upgraded  Molonglo Observatory Synthesis Telescope \citep[MOST;][]{2017bailes} have also proven successful. However, the concomitant increase in the computing cost of forming and searching hundreds of beams in real time have required new approaches to data handling.

MeerTRAP (\textit{More} TRAnsients and Pulsars) is such a survey with the MeerKAT telescope. Further details about this project can be found in \citet{2018sanidas}, \citet{2020jankowski}, \citet{2020malenta}, \citet{2021rajwade}, and a forthcoming system description paper (Stappers et al.\ in prep.). MeerTRAP operates on a commensal basis, meaning that during any given MeerKAT observation it can receive the digitised and channelised data from hundreds of coherent beams formed on the sky using a custom beamformer \citep[][]{2021chen}, and search those beams in real time for single pulses. Because of the limiting cost of data storage, all these data can not be stored permanently, and the single pulse search must take place in real time. If the custom single pulse search algorithm is triggered then the beamformed dynamic spectra for the candidate can be saved, which can then be manually inspected to confirm transient detection.

The most prominent classes of Galactic fast transients that MeerTRAP expects to find are radio pulsars, RRATs, and magnetars. Searches for pulsars most often take advantage of their regular repetition, employing either Fourier transforms (FTs) to detect harmonics in the fluctuation power spectra of time series data \citep[e.g.][]{2002ransom}, or Fast Folding Algorithms \citep[FFAs; e.g.][]{1969staelin, 2020morello} to phase-coherently integrate the data folded at trial rotation periods. Since the signal-to-noise ratio (S/N) of a pulsar's time-integrated signal increases by a factor $\sqrt{N}$ with the number of rotations of the pulsar observed $N$, they are generally more likely to be found by these means than by chance observation of single pulses. However, single pulse searches for pulsars are superior to periodicity searches when the pulse energy distribution (PED) is long-tailed \citep{2003mclaughlin}.

Rotating RAdio Transients (RRATs) are believed to be a related population of NSs that differ from canonical radio pulsars only in the regularity of their emission. Although many pulsars emit pulses only intermittently, RRATs are irregular to the degree that they are more likely to be found via single pulse searches than the periodicity searches mentioned above \citep{2011keane}. These sources are often inactive or otherwise undetectable for seconds to years at a time for reasons that remain unclear. More than 120 RRATs have now been identified\footnote{\url{http://astro.phys.wvu.edu/rratalog/}}, most of which were found in pulsar surveys using the Parkes, Arecibo and Green Bank telescopes \citep[e.g. ][]{2006mclaughlin,2015karako-argaman,2009deneva}. At present there is no consensus on the mechanism responsible for RRATs' apparent variability; explanations include bright bursting periods from pulsars that are ordinarily too faint to detect \citep[][]{2006weltevrede}, and occlusion of emission by radiation belts or debris surrounding the NS \citep[][]{2007luo,2006li}.

Magnetars are relatively young\footnote{\url{http://www.physics.mcgill.ca/~pulsar/magnetar/main.html}} (most $<\sim1$~Myr), highly magnetised ($\sim10^{14}$~G) NSs characterised by dramatic variability in their emission---particularly in the X-ray and $\gamma$-ray regime---featuring short, millisecond-duration bursts to major month-long outbursts. In contrast with canonical rotation-powered pulsars,
the bulk of magnetar emission is believed to be powered by the decay of their strong magnetic fields \citep{1995thompson}. Magnetar emission is far more visible at higher energies than is the case for pulsars, with only six of the 31 known magnetars having been detected in the radio frequency band. For the magnetars with confirmed radio pulsations, the radio emission is unusually bright and variable from pulse to pulse \citep[e.g.][]{2017kaspi}, making single pulse searches ideal for finding magnetars in the radio band. The discovery of the magnetar PSR~J1622$-$4950 via its radio pulsations while in a period of X-ray quiescence \citep[][]{2010levin} provides a precedence for finding magnetars through an untargeted radio search. Additionally, renewed interest in magnetars as radio emitters has resulted from the detection of extremely energetic, millisecond-duration radio bursts from the Galactic magnetar SGR~1935+2154, suggesting a link between magnetars and the widely studied, yet still mysterious, extragalactic FRBs \citep[][]{2020bochenek, 2021bailes}. Finding radio sources similar to SGR~1935+2154, which MeerTRAP is well suited to, would be very important for further elucidating the magnetar -- FRB connection. A useful review of magnetars can be found in \citet{2017kaspi}.

In this paper we present the first twelve MeerTRAP Galactic fast transients, their best available localisations, their timing solutions where available, and the results of the follow-up observations with the Lovell and Parkes telescopes. In \S~\ref{sec:observations} we outline the observations that have led to the discoveries as well as follow-up observations, and in \S~\ref{sec:results} we detail the discovered sources and the results of further analysis. A summary follows in \S~\ref{sec:conclusions}.

As a matter of nomenclature, those sources from which we have seen repeat pulses are referred to by conventional pulsar names consisting of the PSR- prefix and their J2000.0 celestial coordinates. Sources that have been detected only once have not been given such a name, and are referred to using the prefix MTP- and a number indicating the order of the sources' discovery. An exception is MTP0008/PSR~J1901+0254, which was only identified once in a MeerKAT IB, but which was later confirmed to be a pulsar first discovered in the Parkes multibeam pulsar survey \citep[PMPS;][]{2004hobbs}.

\section{Observations and data processing}
\label{sec:observations}

MeerKAT observing is scheduled so that a large fraction of time is spent on the MeerKAT Large Survey Projects (LSPs). Since becoming fully operational in September of 2020, MeerTRAP has been collecting as much commensal data as has been possible given technical and organisational limitations. Thus far many of these observations have taken place during both the initial MeerKAT testing time, and especially that undertaken by the LSP concerned with pulsar timing, viz.~MeerTime \citep[][]{2020bailes}. The vast majority of MeerTime's observations have been of pulsars within a few degrees of the Galactic plane, and as such the commensal data collected by MeerTRAP were initially heavily skewed towards the Galactic plane. However, this is no longer the case as other, more extragalactically-focused LSPs have begun their observing campaigns. Thus, the sources presented in this paper should not be considered especially representative of their populations or of those MeerTRAP will discover in the future. A full population study of MeerTRAP sources will follow once more observing time has accrued.

A full description of MeerTRAP's real-time data processing system will be presented in a forthcoming publication (Stappers et al.\ in prep.), while the single pulse search algorithm is delineated in \citet{2020malenta} and a preliminary overview is given by \citet{2021rajwade}. In regular operation, the on-site Filterbank BeamFormer User Supplied Equipment (FBFUSE) cluster, developed by collaborators at the Max-Planck Institut f\"ur Radioastronomie \citep[][]{2021chen}, uses channelised complex voltage data from the MeerKAT UHF-band (544 -- 1088~MHz) or L-band receivers (856 -- 1712~MHz) to form up to 768 coherent beams (CBs) on the sky, typically using about 40 dishes in the dense core region of the MeerKAT telescope array. An incoherent beam (IB) is also formed using the sum of the data from up to 64 MeerKAT antennas. The data are taken with a sampling time of approximately 482~$\mu$s in the UHF band and approximately 306~$\mu$s in the L-band. The beamformed data are transferred over the network to MeerTRAP's computer cluster, Transient User Supplied Equipment (TUSE), consisting of 66 compute servers, one of which serves as a head node that manages the other nodes. Each compute node is equipped with two 16-core Intel Xeon processors, two Nvidia GTX 1080 Ti Graphical Processor Units (GPUs) with 12 GB memory, and 256 GB of RAM, and is capable of processing up to 12 beams worth of data at once. On each node the data are transposed into frequency-time format and committed to shared memory ring buffers.

In order to excise Radio Frequency Interference (RFI), we have applied a static frequency channel mask to remove channels known to be affected. Additionally, we apply standard zero-DM excision \citep{2009eatough}. The cleaned data are then fed into an augmented version of the real-time, GPU-based single pulse search software \texttt{ASTROACCELERATE} described in \citet{2019carels}. This involves incoherently dedispersing the time series for each beam to various trial DMs of up to 5000~pc~cm$^{-3}$, and convolving the result with a series of boxcar filters of varying widths up to 0.67~s. Any response with a S/N above eight is taken as a candidate of interest to be further appraised. This is done first using a machine learning algorithm based on the \texttt{FETCH} (Fast Extragalactic Transient Candidate Hunter) classifier \citep[][]{2020agarwal} with a probability threshold of greater than 0.5. Surviving candidates are then evaluated by eye. Frequency-time data for candidates of interest are written and saved in the commonly-used \texttt{SIGPROC}\footnote{\url{http://sigproc.sourceforge.net/}} \texttt{filterbank} format \citep[][]{2011lorimer}.

During regular MeerTRAP operation, the CBs are tiled to overlap at 25 per cent of their maximum sensitivity\footnote{Whenever in this paper an overlap percentage is cited, it is meant in the sense that the CBs are arranged so that they intersect at this proportion of their maximum sensitivities at the centre frequency. This is determined by an ellipse fit to a model of the CBs' PSF at $\sim$1284~MHz. For more details on how this is achieved see \cite{2021chen}.}, which represents a compromise between the total CB FoV and the sensitivity between CBs. A single pulse would normally only be detected if it came from within the extent of this tiling. However, more recently we have implemented a scheme whereby, when operating commensally with an observation that happened to be within view of a confirmed MeerTRAP candidate, we tile 19 CBs on the candidate coordinates in addition to the regular beams around the target source position. These 19 beams are arranged to overlap at 84 per cent of their maxima (i.e.~Nyquist sampled), so as to retain sufficient spacing between CBs to ensure the source is inside the tiling pattern, while maximising the number of CBs a pulse would be detected in at once for the greatest potential of CB localisation. As mentioned above, since the preponderance of early MeerTRAP observations were towards the Galactic plane, sources in the plane were more likely to have been discovered by MeerTRAP earlier than those far from the plane, as well as to have had more MeerKAT observations. The exception is PSR~J0930$-$1854 (MTP0007), which has had the overall most MeerKAT observing time because it is situated within the field of PSR~J0931$-$1902, a regular target of MeerTime\footnote{\url{https://pulsars.org.au/public/J0931-1902}}.

\begin{figure}
    \centering
    \includegraphics[width=.48\textwidth]{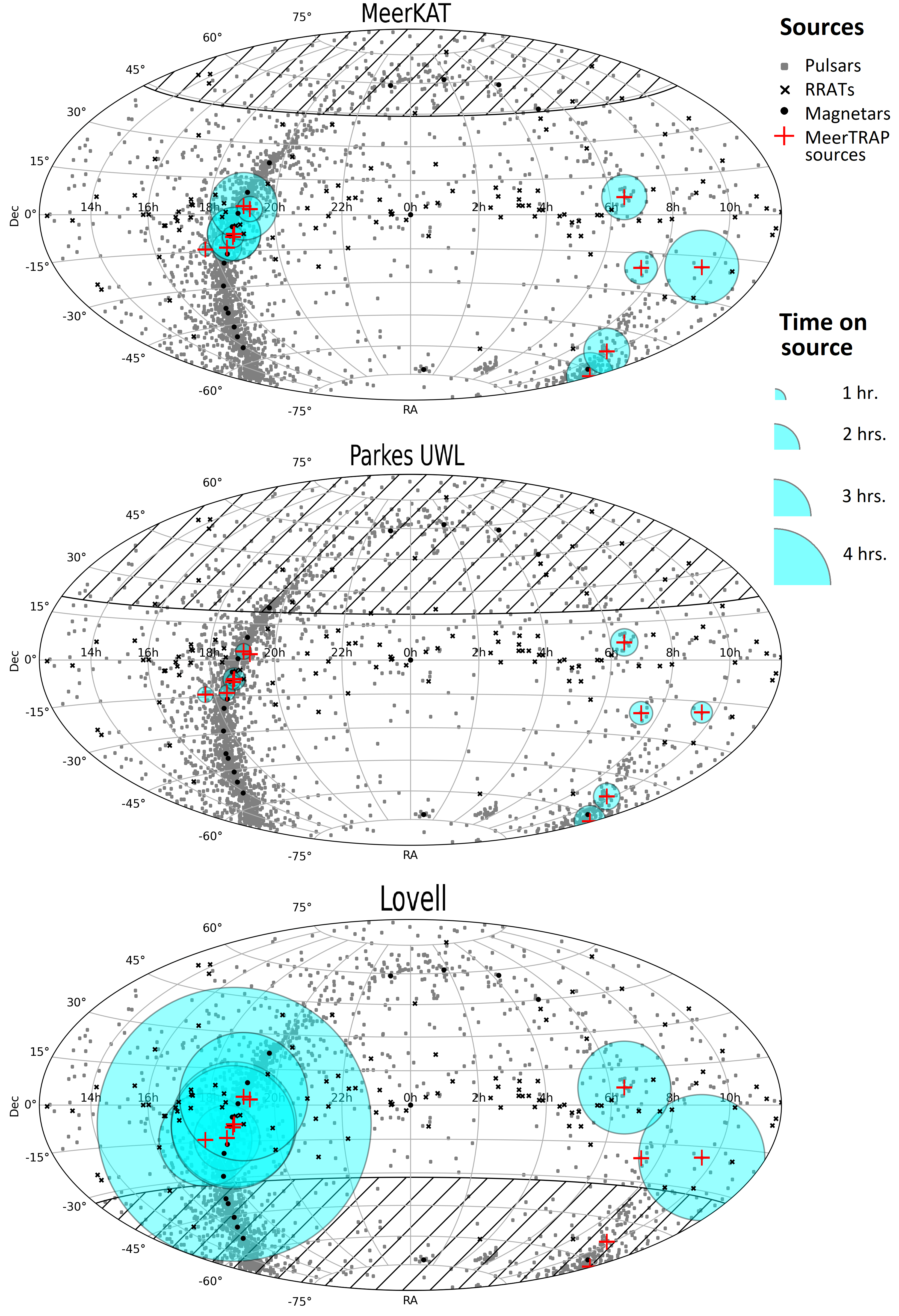}
    \caption{Aitoff projection showing the location of known pulsars (grey dots), RRATs (black crosses), and magnetars (black dots), as well as the twelve Galactic MeerTRAP sources indicated by red crosses. The size of the cyan circles centred on the MeerTRAP sources' coordinates indicates the amount of time spent observing each source's coordinates with the MeerKAT (top panel), Parkes (middle), and Lovell telescopes, respectively. The hatching shows the coordinates beyond the approximate declination limit of each telescope.}
    \label{fig:ObsTimes}
\end{figure}

Following confirmation of a candidate pulse as genuine, it is also earmarked for follow-up observations with various instruments. Because MeerTRAP is a commensal survey project, we are limited in our ability to perform follow-up observations of the sources that we discover. As we are not able to dictate MeerKAT's observing schedule, we followed up on our MeerKAT detections with observations using both the Lovell and the Parkes radio telescopes.

We used the ultra-wide-bandwidth low-frequency (UWL) receiver on the 64-m Parkes radio telescope to observe the MeerTRAP Galactic sources. The UWL receiver operates over a continuous frequency band from 704 to 4032~MHz, divided into 26$\times$128~MHz sub-bands which are amplified separately \citep[][]{2020hobbs}. The sub-band voltage data were processed using the GPU-based `Medusa' signal processor system in pulsar searching mode, which uses the \texttt{DSPSR} \citep{2011vanstraten} package to produce data files in the \texttt{PSRFITS} \citep{2004Hotan} format. The data were taken with a sampling time of $512~\mu \textnormal{s}$, 8-bit resolution, and 3328 frequency channels across the bandwidth. For each source the best known DM measured by MeerTRAP was used to coherently dedisperse the data.

We performed observations of MeerTRAP Galactic sources with the Parkes telescope on eleven occasions between 31 May 2020 and 19 March 2021, amounting to almost twelve hours across eleven of the twelve sources. The amount of time spent on each source was constrained by the local sidereal time (LST) of the awarded observing time, so that we were able to gather around two hours' worth of PSR~J1152$-$6056 (MTP0002), but only 20 minutes on PSR~J1840$-$0840 (MTP0005), and no time at all on MTP0012, which was only identified as a discovery towards the end of the follow-up campaign.

We have searched all the UWL data for single pulses from the sources. The data were first cleaned using a static channel mask as well as the \texttt{CLFD}\footnote{\url{https://github.com/v-morello/clfd}} \citep[][]{2019morello} and/or \texttt{IQRM}\footnote{\url{https://github.com/v-morello/iqrm}}\citep{2021morello} algorithms. We then produced single pulse stacks using \texttt{DSPSR}. For PSR~J1843$-$0757 (MTP0001), PSR~J1152$-$6056 (MTP0002), and PSR~J0943$-$5305 (MTP0011) we folded the data using their established timing solutions; the rest were folded at a nominal period and inspected manually to look for single pulses. It should be noted that the UWL data were often heavily affected by RFI which, despite the RFI mitigation methods above, may have significantly impaired our ability to detect single pulses. 

There have been Lovell telescope observations of eight of the sources, that is, all but PSR~J1152$-$6056 (MTP0002) and PSR~J0943$-$5305 (MTP0011), which fall below the telescope's declination limit, and PSR~J0723$-$2050 (MTP0010) and MTP0012, which were identified too late to enter into the observing roster. The observations were processed using the \texttt{ROACH} backend\footnote{Reconfigurable Open Architecture Computing Hardware (ROACH) FPGA board developed by the Collaboration for Astronomy Signal Processing and Electronics Research (CASPER) group; \url{http://casper.berkeley.edu/}}. The ROACH applies a polyphase filter to channelise data from the 1332-1772~MHz frequency range into 25$\times$16~MHz subbands. Using \texttt{DSPSR} each subband is then further channelised into 32$\times$0.5~MHz channels and downsampled to a sampling time of 256~$\mu$s. We then followed the same RFI excision and folding procedure as described above for the Parkes UWL data.

It is important to note that the Lovell observations were also heavily affected by RFI\footnote{Observations of MeerTRAP sources from November 2019 to March 2020 were rendered unusable due to strong interference from a recently installed 4G network transmitter nearby the telescope.}. The Lovell observatory was also inoperative due to the COVID-19 pandemic from March to September 2020, after which improved RFI mitigation greatly ameliorated the data quality. As a result, in the period stretching from September 2020 to June 2021 we have approximately 30 hours worth of usable observations of MeerTRAP sources. As with the UWL observations, we folded the data using established timing solutions or nominal periods to identify possible single pulses.

Figure~\ref{fig:ObsTimes} shows the amount of follow-up time spent on each of the twelve MeerTRAP sources discussed in this paper. The markers in gray show the coordinates of known pulsars recorded in the ATNF pulsar catalog\footnote{\url{http://www.atnf.csiro.au/people/pulsar/psrcat/}} \citep[][]{2005manchester}, RRATs in the RRATalog\footnote{\url{http://astro.phys.wvu.edu/rratalog/}}, and magnetars in the McGill online magnetar catalog\footnote{\url{http://www.physics.mcgill.ca/~pulsar/magnetar/main.html}} \citep[][]{2014olausen}. The red dots show the best-known position of each MTP source, with the size of the overlaid cyan circles corresponding to the number of minutes worth of observing time on each source with MeerKAT (top panel), the Parkes UWL receiver (middle), and the Lovell telescope (bottom). The hatched areas show the declination limits beyond which each instrument cannot observe, i.e.~$\leq$+44$^\circ$ for MeerKAT \citep[][]{2018camilo}, $\leq$+20$^\circ$ for Parkes \citep[][]{2010keith}, and $\geq$$-$32$^\circ$ for the Lovell \citep[][]{2010keane}.

\section{Results}
\label{sec:results}

\begin{figure*}
    \centering 
        \begin{subfigure}[b]{0.33\textwidth}
            \centering
            \includegraphics[width=\textwidth]{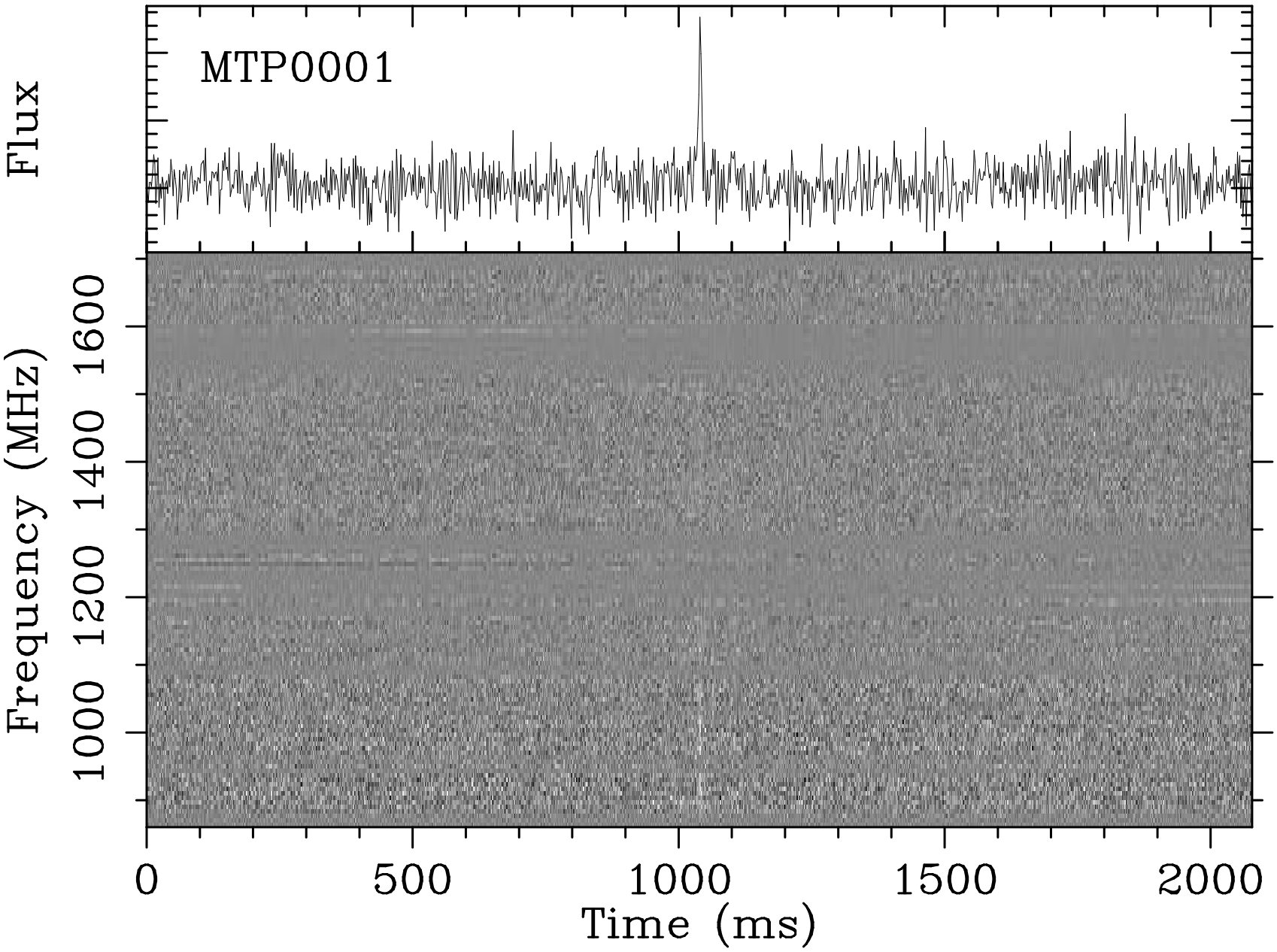}
            \label{fig:MTP1}
        \end{subfigure}
        \begin{subfigure}[b]{0.33\textwidth}  
            \centering 
            \includegraphics[width=\textwidth]{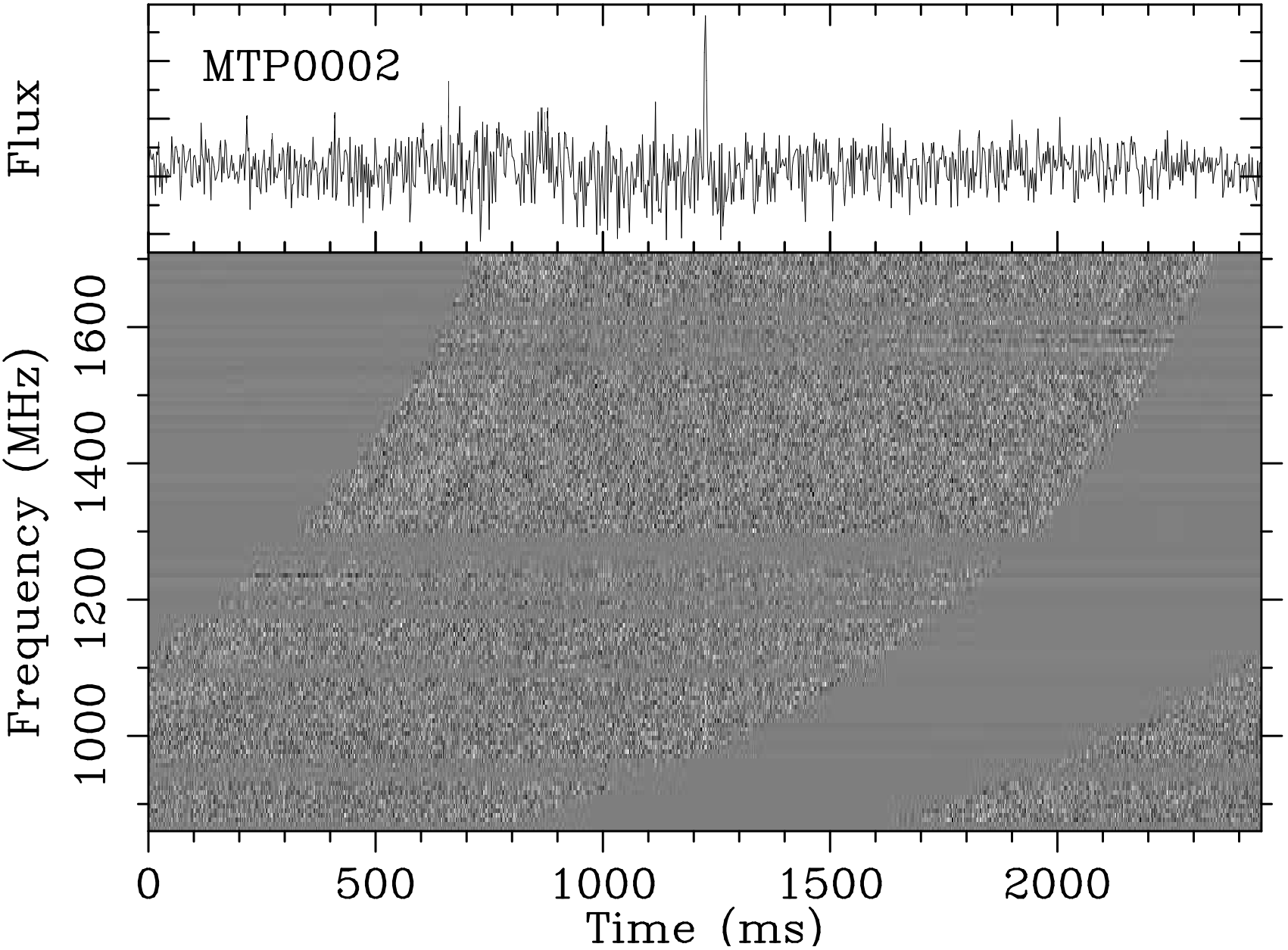}
            \label{fig:MTP2}
        \end{subfigure}
        \begin{subfigure}[b]{0.33\textwidth}   
            \centering 
            \includegraphics[width=\textwidth]{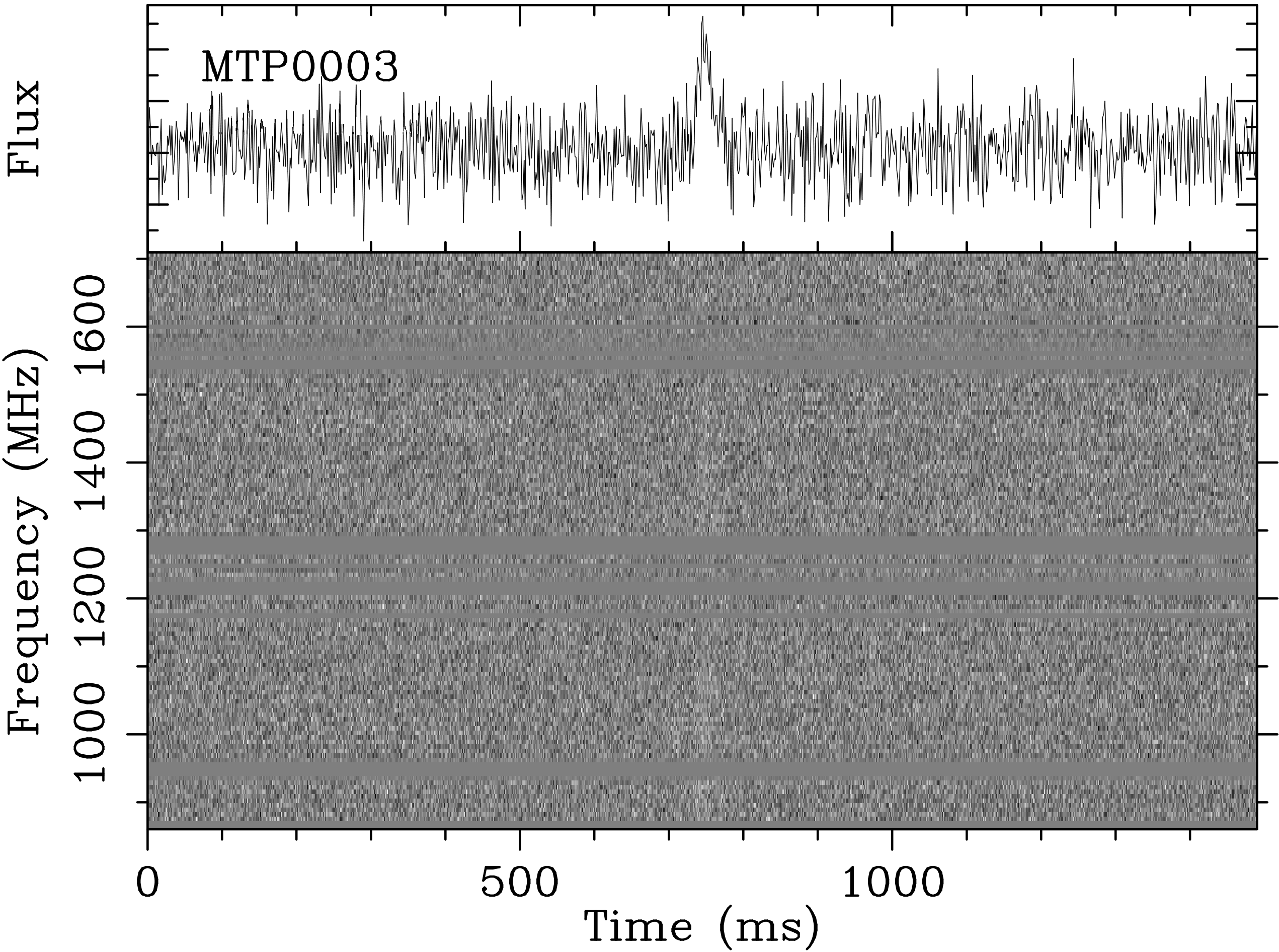}
            \label{fig:MTP3}
        \end{subfigure}
        \hfill
        \begin{subfigure}[b]{0.33\textwidth}   
            \centering 
            \includegraphics[width=\textwidth]{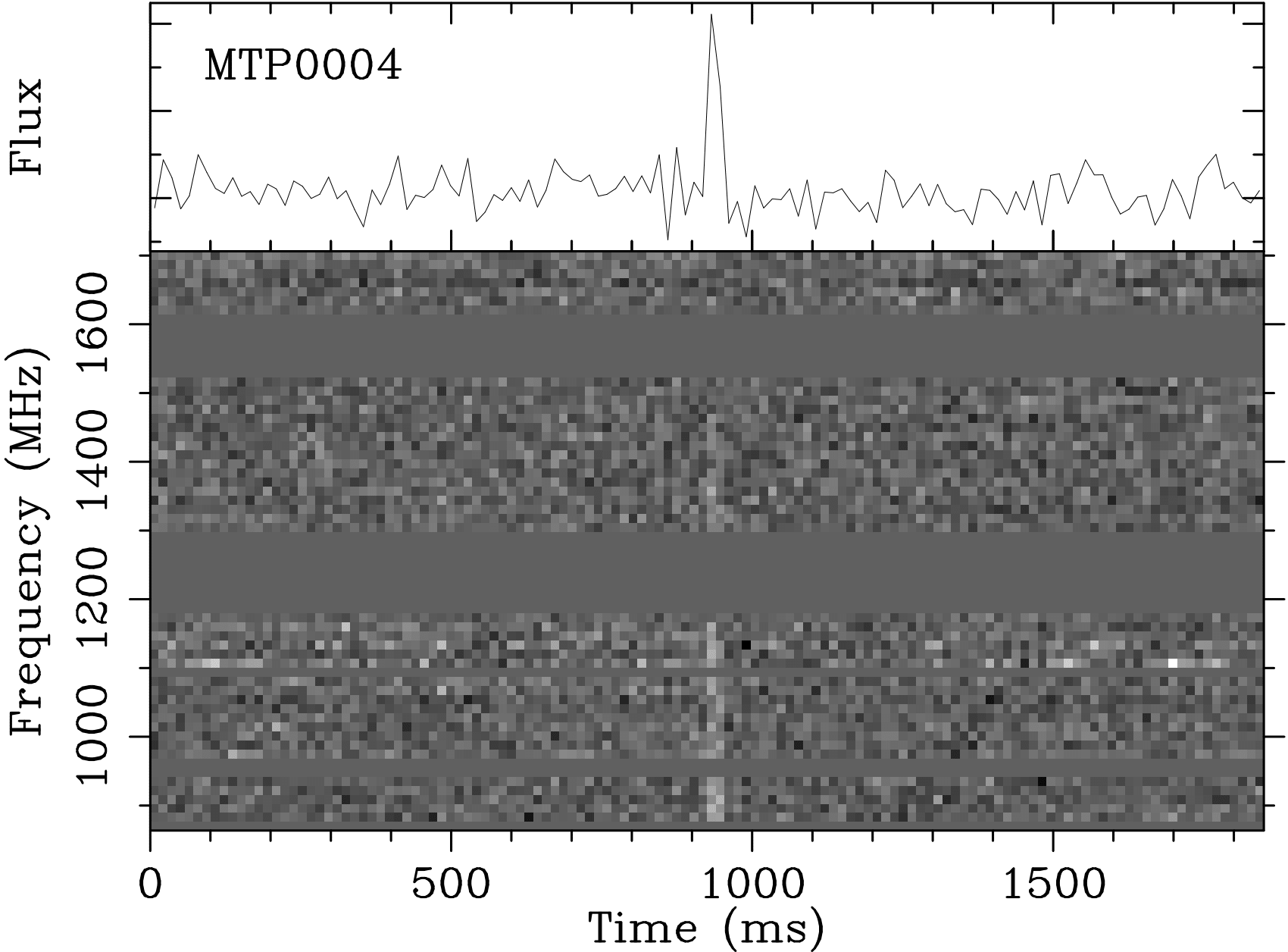}
            \label{fig:MTP4}
        \end{subfigure}
        \begin{subfigure}[b]{0.33\textwidth}
            \centering
            \includegraphics[width=\textwidth]{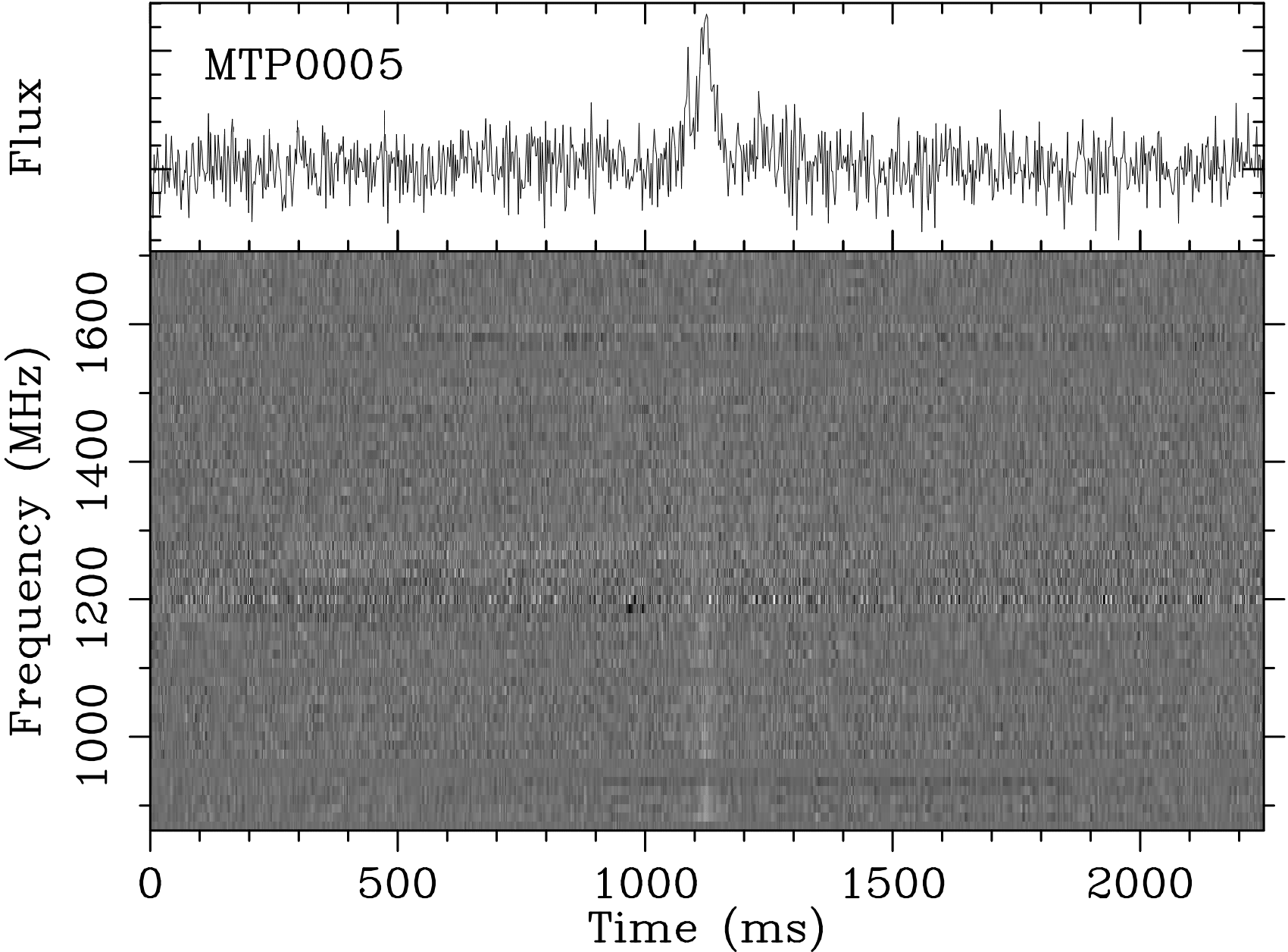}
            \label{fig:MTP5}
        \end{subfigure}
        \hfill
        \begin{subfigure}[b]{0.33\textwidth}  
            \centering 
            \includegraphics[width=\textwidth]{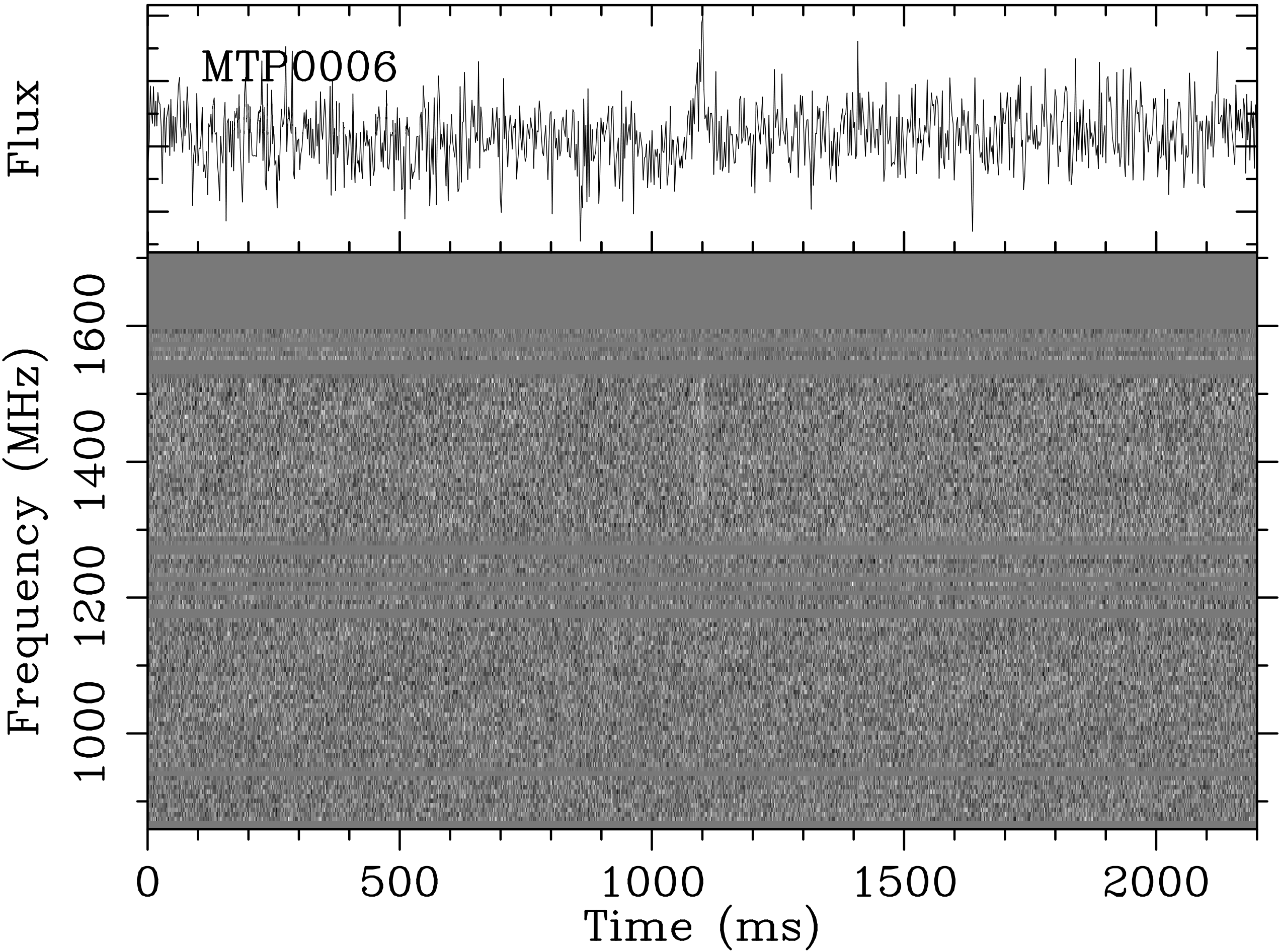}
            \label{fig:MTP6}
        \end{subfigure}
        \begin{subfigure}[b]{0.33\textwidth}
            \centering
            \includegraphics[width=\textwidth]{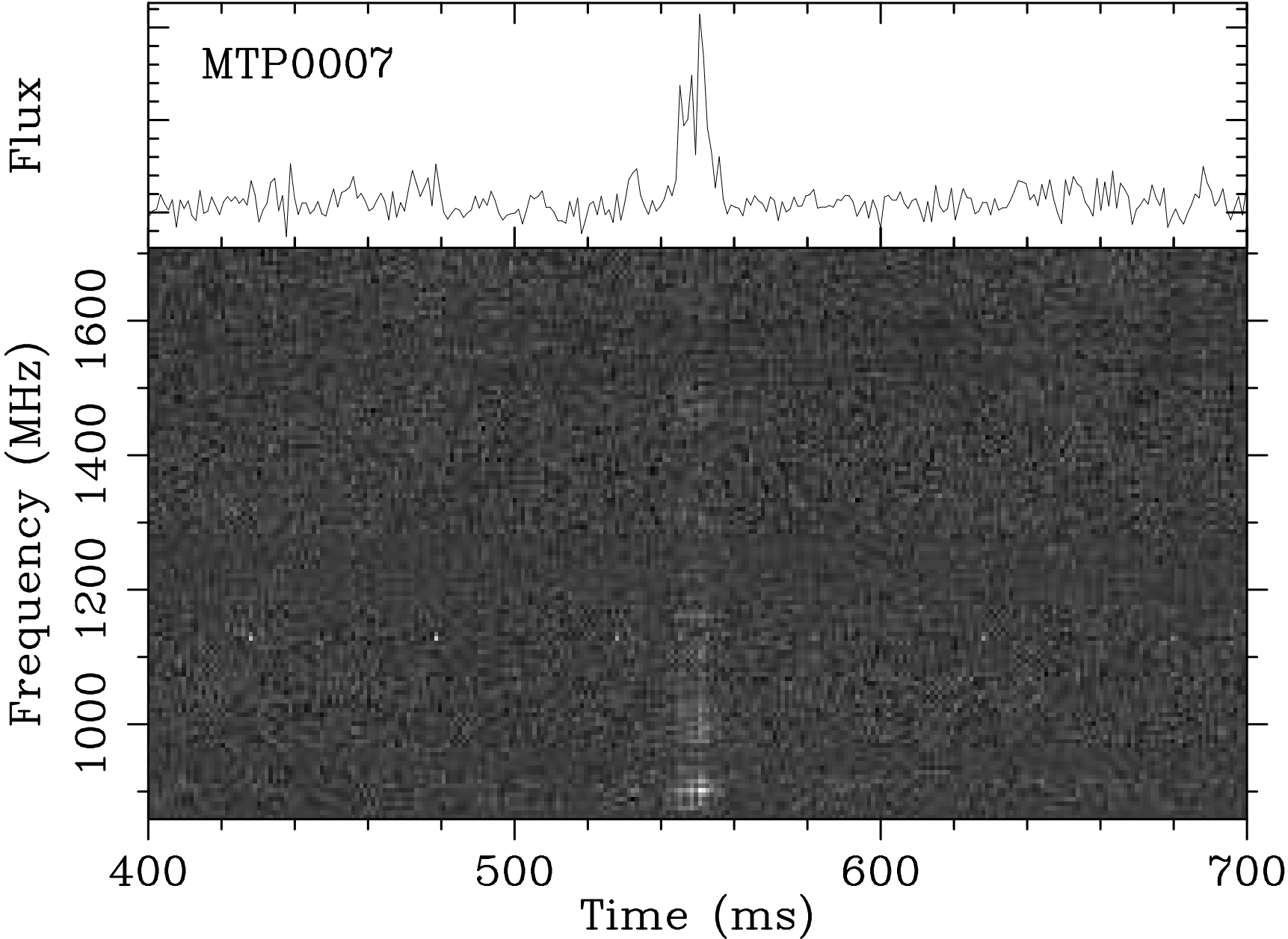}
            \label{fig:MTP7}
        \end{subfigure}
        \hfill
        \begin{subfigure}[b]{0.33\textwidth}  
            \centering 
            \includegraphics[width=\textwidth]{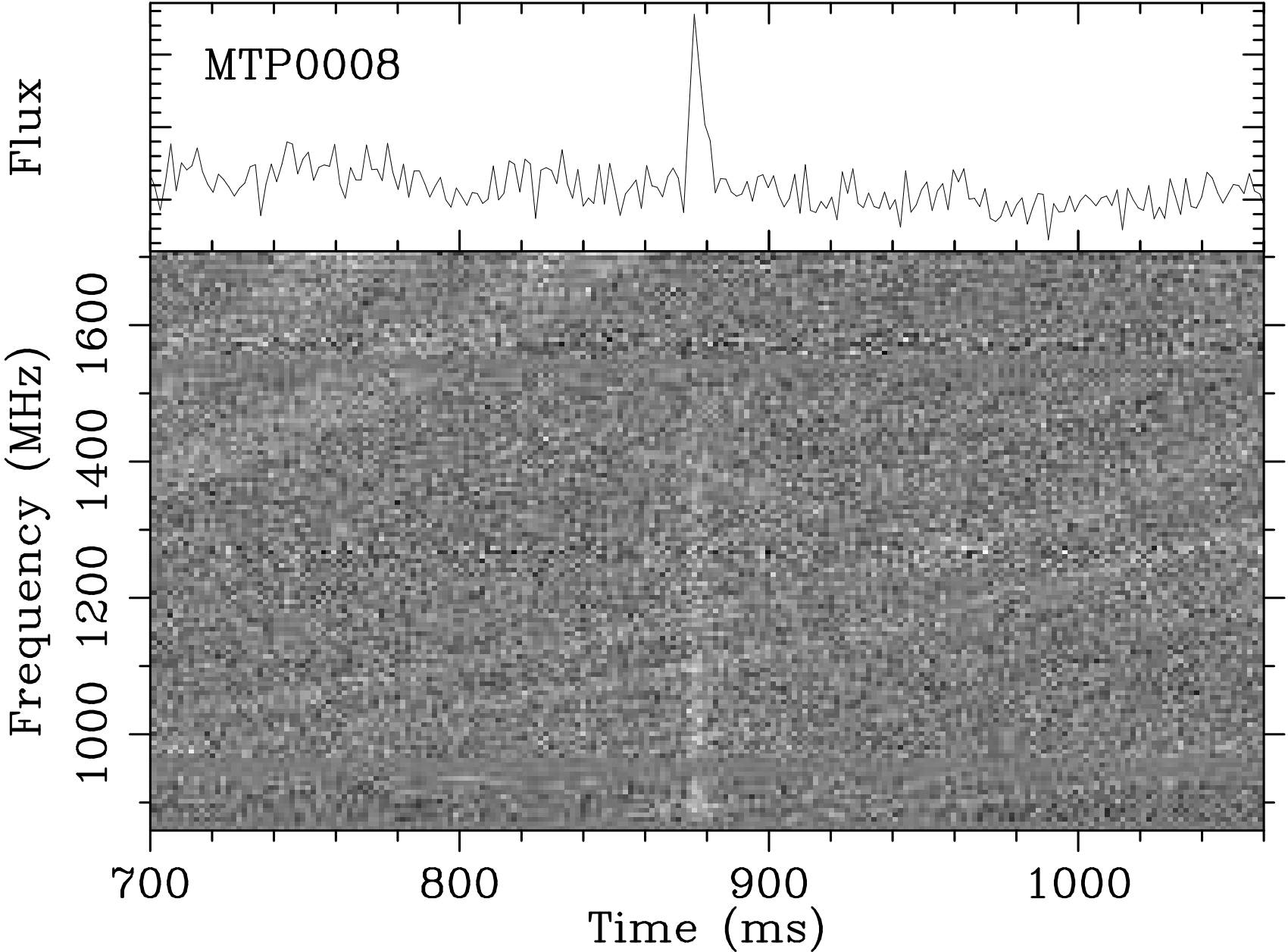}
            \label{fig:MTP8}
        \end{subfigure}
        \begin{subfigure}[b]{0.33\textwidth}
            \centering
            \includegraphics[width=\textwidth]{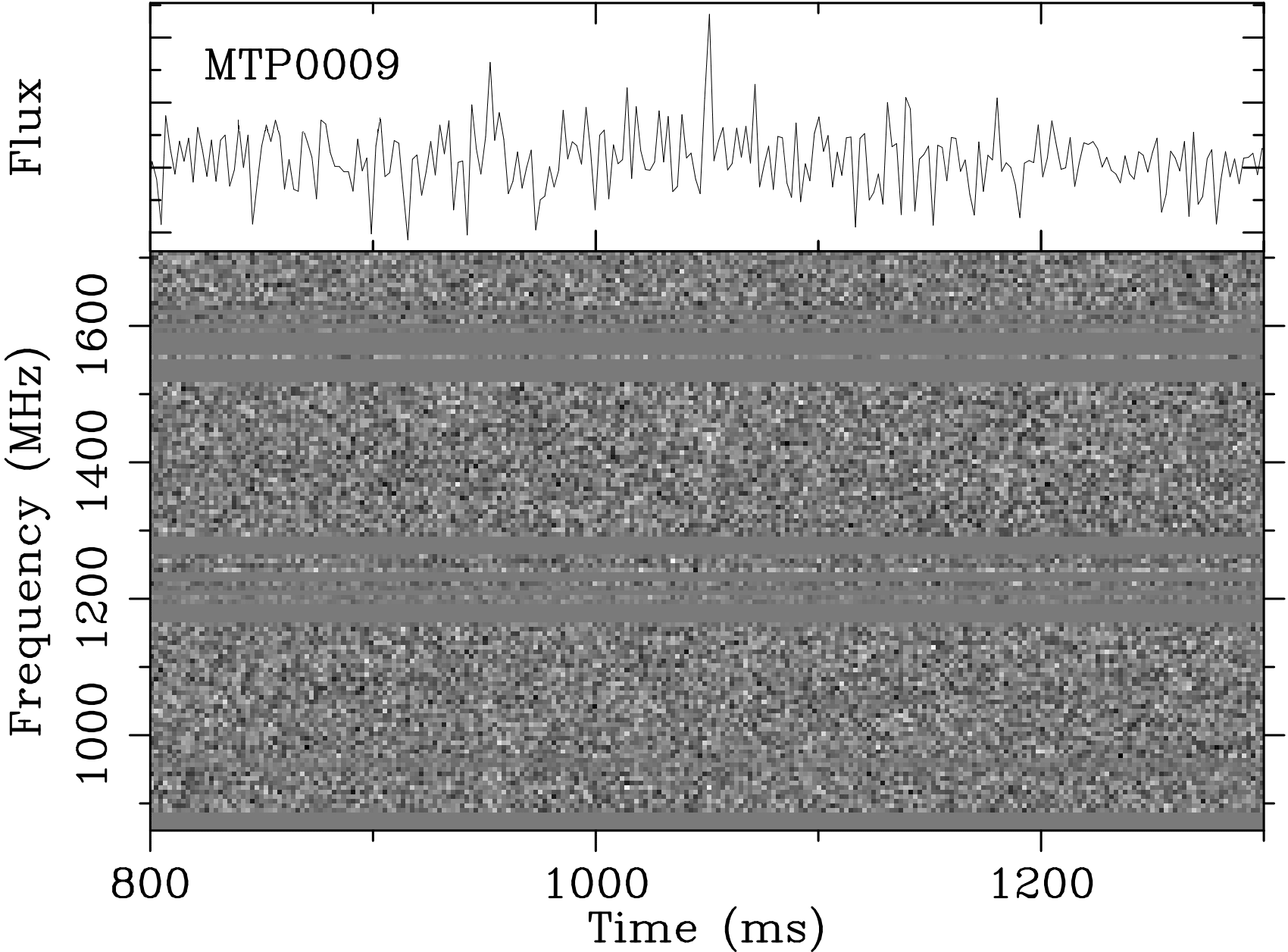}
            \label{fig:MTP9}
        \end{subfigure}
        \hfill
        \begin{subfigure}[b]{0.33\textwidth}  
            \centering 
            \includegraphics[width=\textwidth]{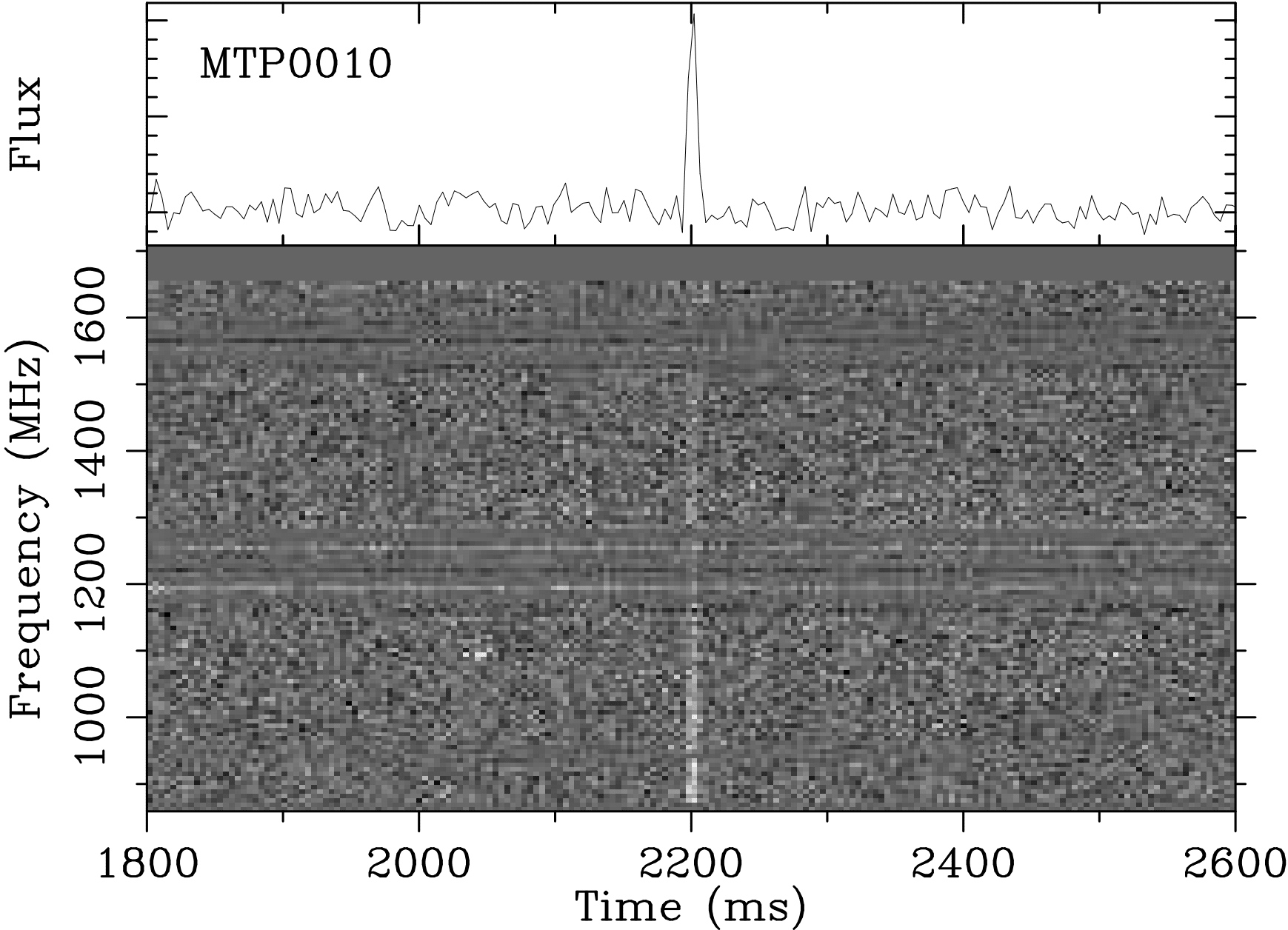}
            \label{fig:MTP10}
        \end{subfigure}
        \begin{subfigure}[b]{0.33\textwidth}
            \centering
            \includegraphics[width=\textwidth]{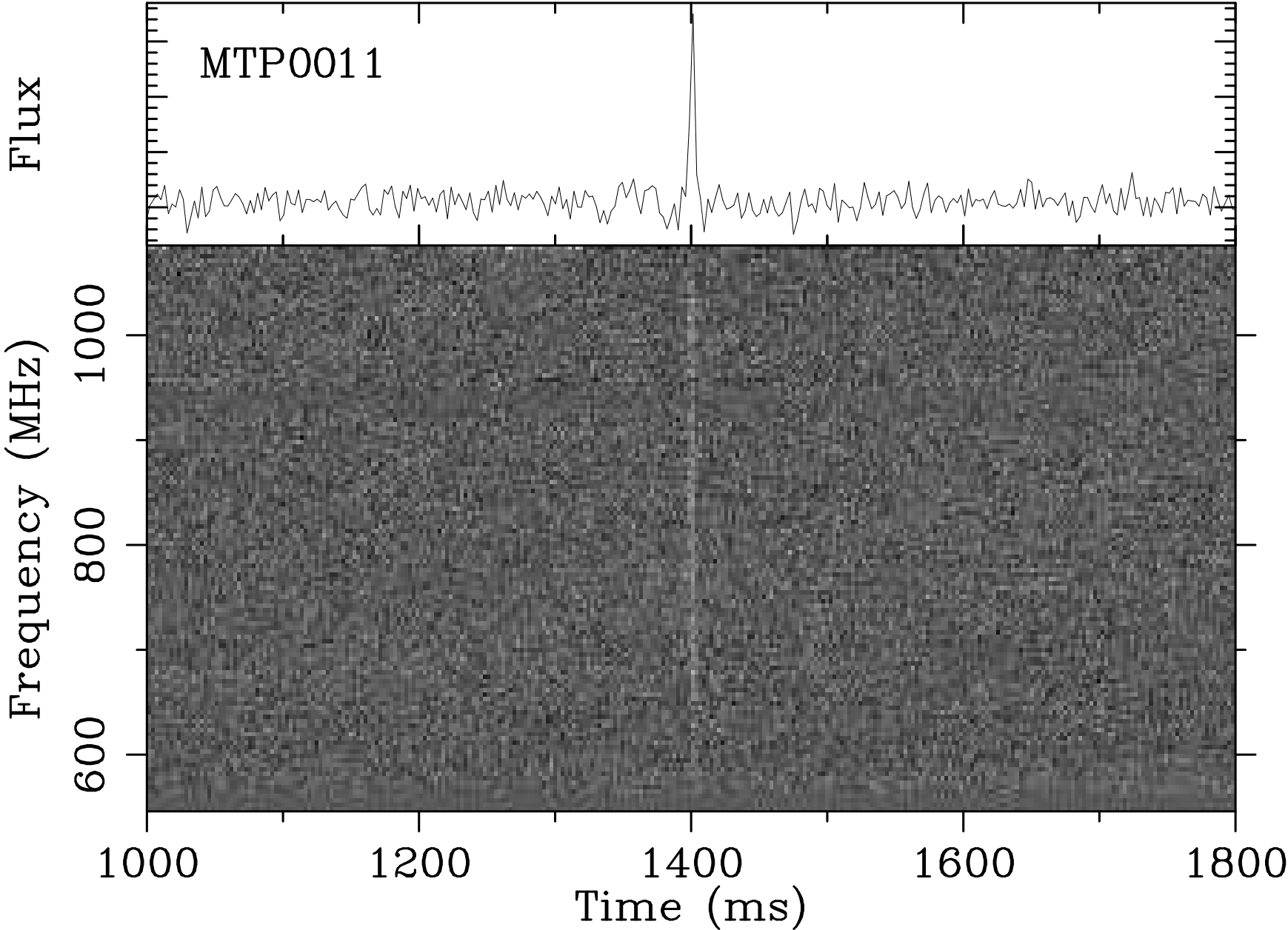}
            \label{fig:MTP11}
        \end{subfigure}
        \hfill
        \begin{subfigure}[b]{0.33\textwidth}  
            \centering 
            \includegraphics[width=\textwidth]{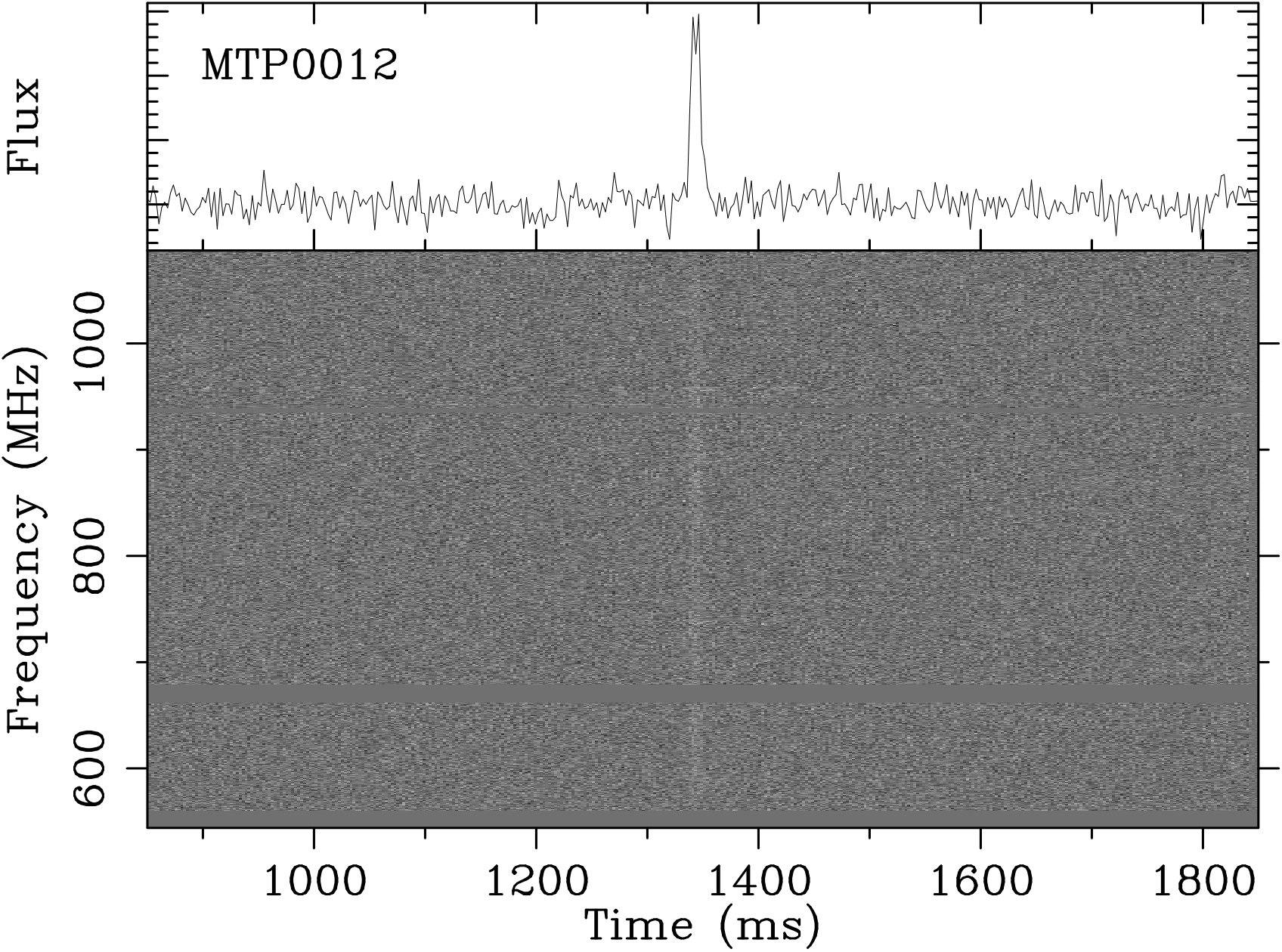}
            \label{fig:MTP12}
        \end{subfigure}
    \caption{Dynamic spectra (bottom) and pulse profile (top) for the discovery pulses of each of the twelve Galactic MeerTRAP sources. The sources are arranged from left-to-right and top-to-bottom in the order of discovery. Note that all of the discoveries were made using the MeerKAT L-band receiver (856 -- 1712~MHz), except for PSR~J0943$-$5305 and MTP0012, which were found in the UHF band (544 -- 1088~MHz). The sweeping grey patches in the MTP0002/PSR~J1152$-$6056 plot represent missing data, and are present because the filterbank data saved for the discovery pulse were not long enough to include the full dispersive delay of the soure. Conversely, the sweeping features in the MTP0008/PSR~J1901+0254 plot are due to zero-DM RFI.}
    \label{fig:profiles}
\end{figure*}

A total of twelve new Galactic sources have been confirmed via detection of at least one pulse identified above a signal-to-noise limit of eight by the MeerTRAP single pulse detection pipeline and evaluated by eye. Table~\ref{table:initial_parameters} contains a list of these sources and the attributes of the discovery pulse. Note that the listed parameters are only for the discovery pulse(s); certain sources that were discovered via a single pulse were assigned a PSR- name based on later repeat pulses.

\begin{table*}
\caption{Discovery parameters of the twelve new Galactic fast transient sources found by MeerTRAP. Only sources from which we have seen repeat pulses are given PSR- names. Sources marked by asterisks were discovered in the IB of MeerKAT only. For sources from which more than one pulse was detected at discovery, the listed coordinates are those of the beam where the brightest pulse was detected; the pulse width and S/N for these sources are also recorded as a range of values.}\label{table:initial_parameters}
\begin{tabular}{lcrrrrrrr}
\toprule
Source   & PSR             & RA (hms) & Dec (dms) & DM (pc~cm$^{-3}$) & \begin{tabular}[c]{@{}l@{}} DM$_{\mathrm{Gal}}$ (pc~cm$^{-3}$)\\ \textsc{ne2001}/\textsc{ymw16}\end{tabular} & \begin{tabular}[c]{@{}l@{}}DM distance (kpc)\\ \textsc{ne2001}/\textsc{ymw16}\end{tabular} & Pulse width (ms) & S/N     \\
\midrule
MTP0001  & J1843$-$0757 & 18:43:31 & $-$07:57:22 & 254    &   893 / 595  & 4.9 / 5.0                                                               & 4.6              & 12      \\
MTP0002  & J1152$-$6056 & 11:52:37 & $-$60:56:22 & 381    &  654 / 828   & 7.8 / 6.4                                                               & 6.1              & 9       \\
MTP0003  &                  & 17:42:35 & $-$14:01:39 & 124     & 289 / 207   & 3.2 / 4.4                                                               & 18.4             & 8       \\
MTP0004 &                  & 18:24:52 & $-$13:31:19 & 212      & 1653 / 2297  & 3.7 /3.4                                                                & 5.8              & 14      \\
MTP0005*  & J1840$-$0840 & 18:42:55 & $-$08:00:54 & 299      & 922 / 621  & 5.4 / 6.2                                                               & 15.3 - 27.0      & 10 - 12 \\
MTP0006  &                  & 18:39:51 & $-$09:10:00 & 295 -- 304 & 1008 / 674  & 4.9 / 5.8                                                               & 4.6 - 7.7        & 8       \\
MTP0007  & J0930$-$1854 & 09:30:34 & $-$18:54:54 & 33    &  68 / 53  & 1.5 / 2.5                                                               & 1.5              & 9       \\
MTP0008* & J1901+0254   & 19:03:06 & 03:27:19  & 181    &  913 / 770   & 5.0 / 4.5                                                               & 3.7              & 8       \\
MTP0009  &                  & 06:27:15 & 06:56:27  & 262     &  184 / 304  & $>$42.1 / 7.0                                                               & 2.1              & 8       \\
MTP0010  & J0723$-$2050 & 07:23:02 & $-$20:50:04 & 130    &  198 / 332   & 4.8 / 3.3                                                               & 3.7              & 8 - 12  \\
MTP0011  & J0943$-$5305 & 09:43:08 & $-$53:05:17 & 174     &  456 / 632  & 3.9 / 0.4                                                               & 8.7              & 9 - 21  \\
MTP0012  &   & 19:14:43 & 02:18:34  & 161    &  442 / 270   & 5.0 / 7.4                                                               & 9.2              & 21     \\
\bottomrule
\end{tabular}
\end{table*}

Figure~\ref{fig:profiles} compiles plots of the discovery pulses from each of the sources. In each, the bottom panel shows the intensity of the pulse as a function of frequency and time, and the top panel shows a frequency averaged profile.

\subsection{MTP0001/PSR~J1843$-$0757}

\begin{figure*}
    \centering
    \includegraphics[width=\textwidth]{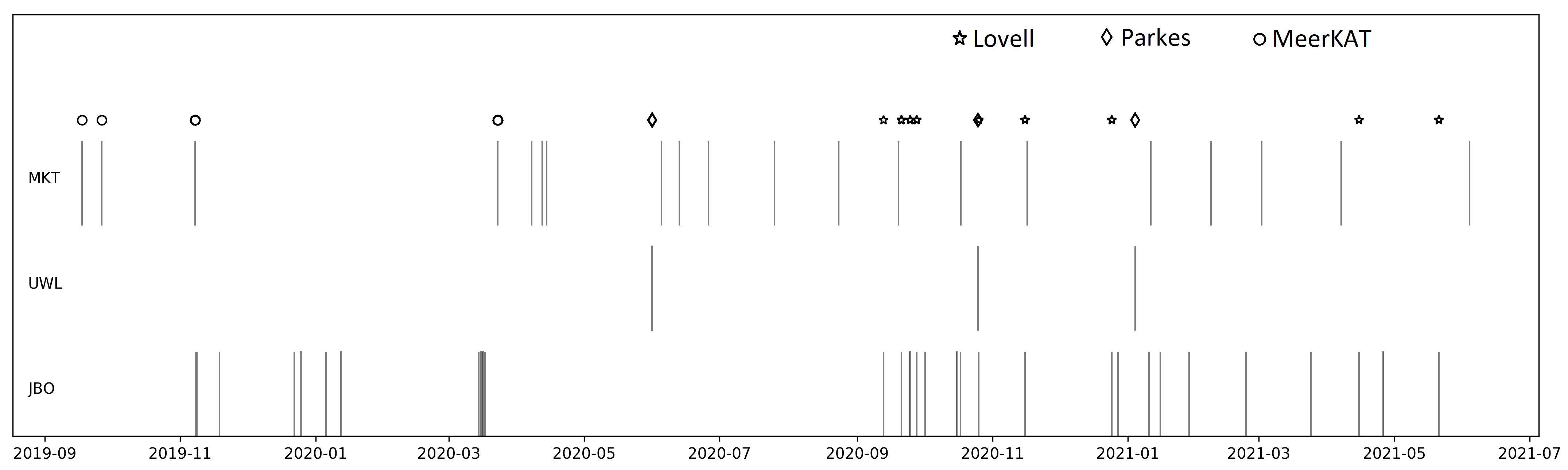}
    \caption{Detection dates for all known pulses of PSR~J1843$-$0757. The markers show days on which at least one pulse was detected, while the vertical lines below indicate periods when the source coordinates were observed using the respective instruments.}
    \label{fig:MTP1_dets}
\end{figure*}

\begin{figure*}
    \centering
    \includegraphics[width=0.32\textwidth]{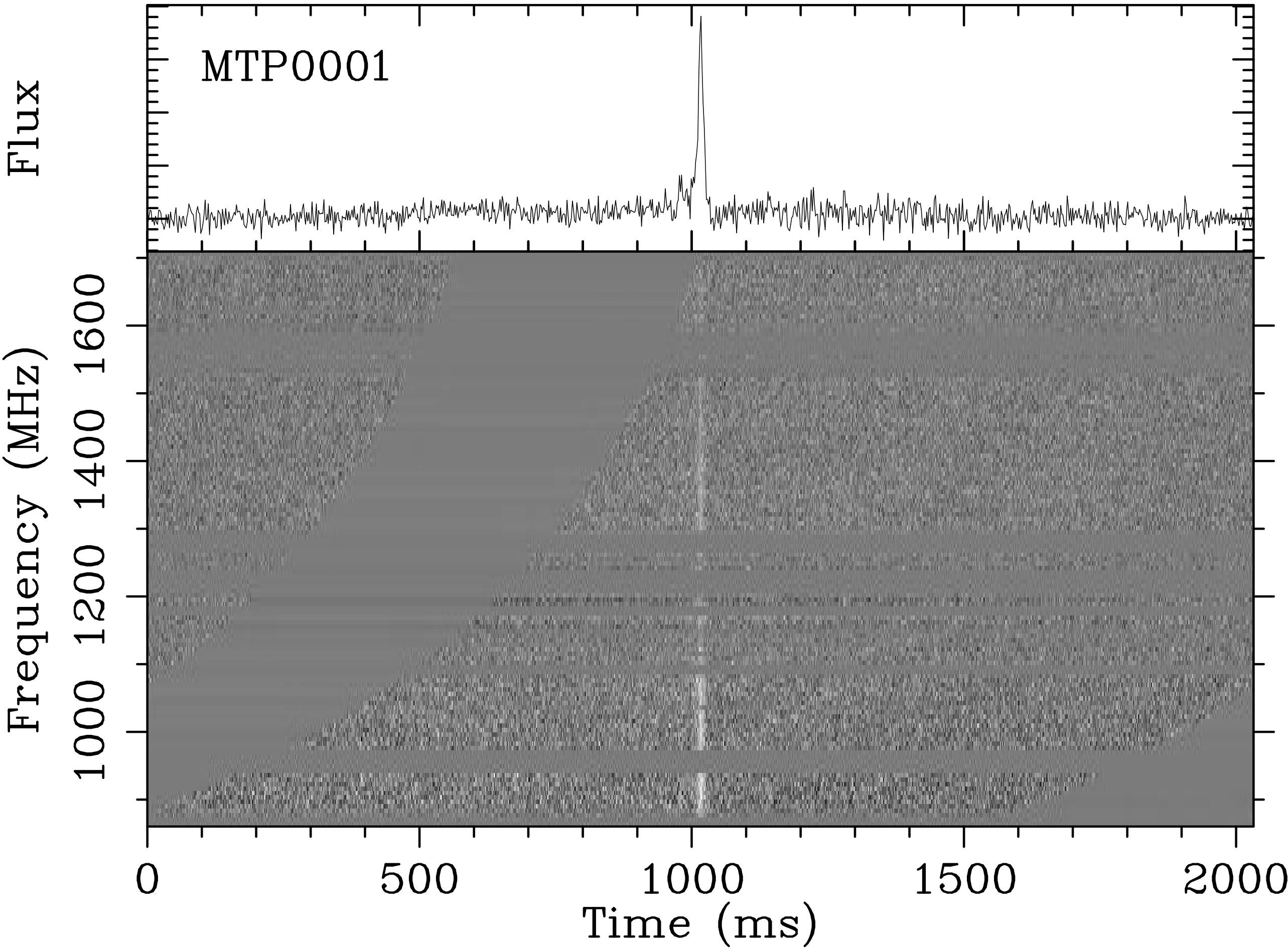}
    \includegraphics[width=0.32\textwidth]{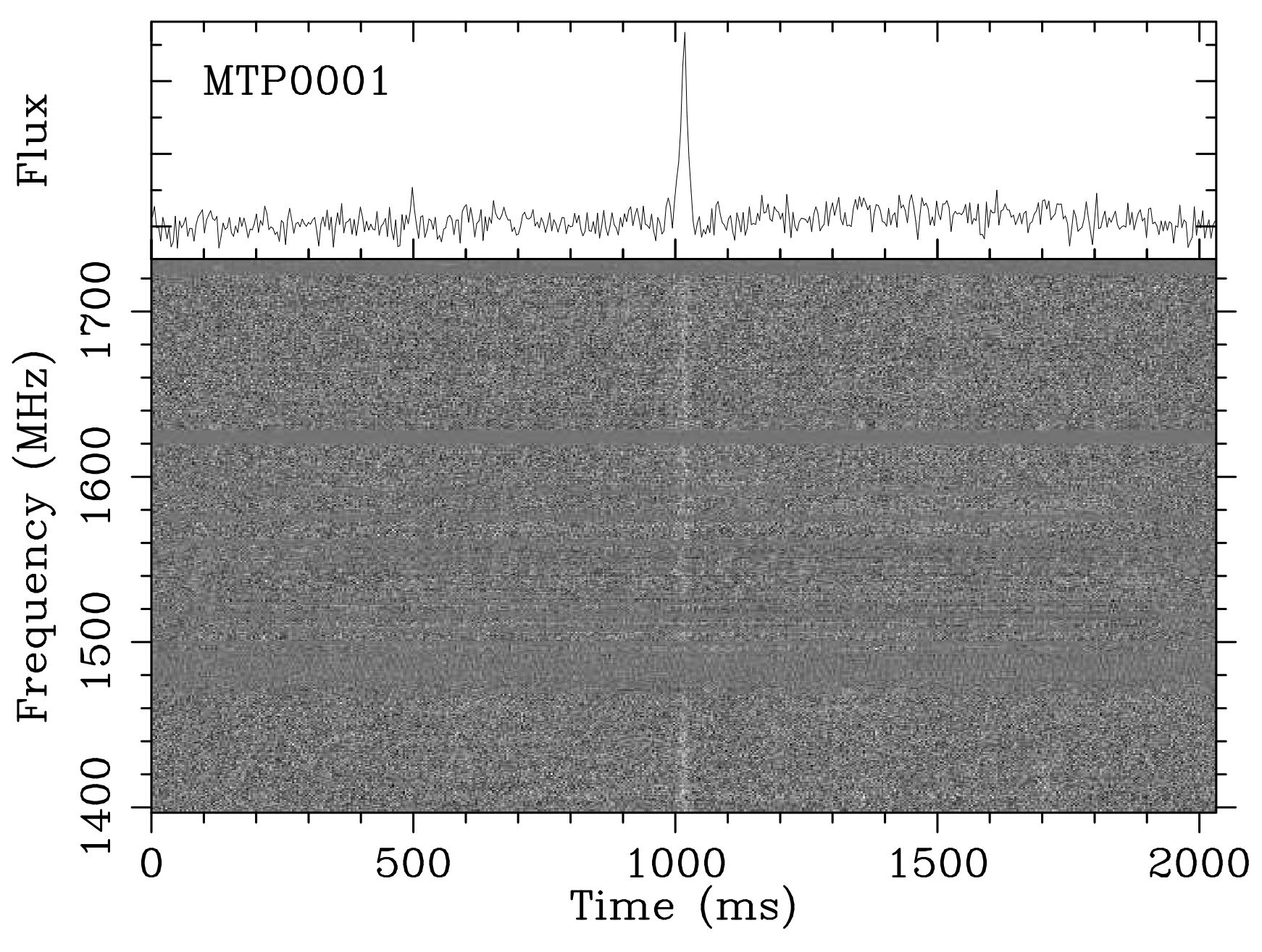}
    \includegraphics[width=0.32\textwidth]{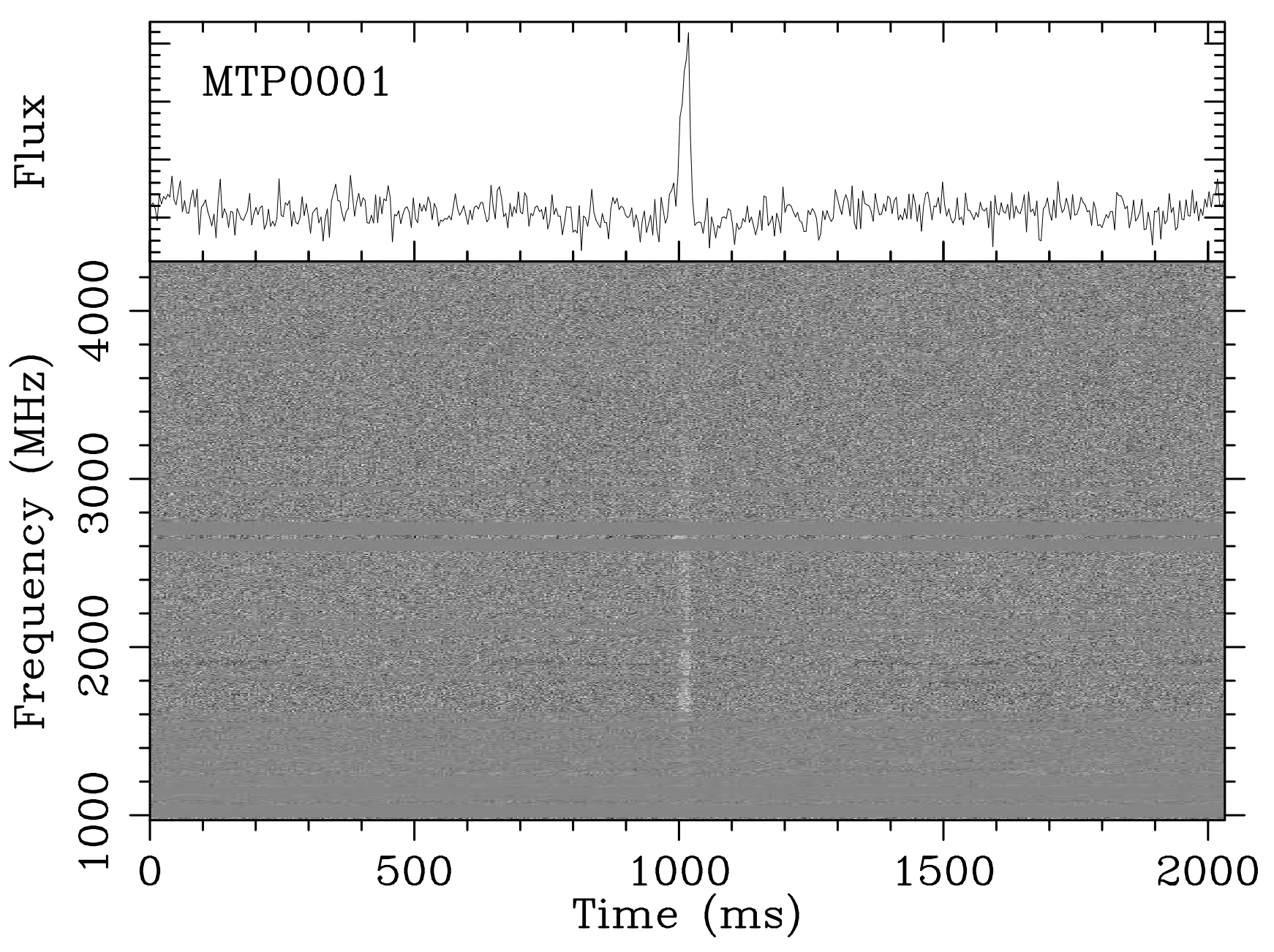}
    \caption{Dynamic spectra (bottom) and profiles (top) of PSR~J1843$-$0757 averaged over 65 MeerKAT detections (left), 51 Lovell detections (middle), and 101 Parkes (right) detections of the source. Note that the sweeping grey patches in the MeerKAT plot represent missing data before and after the dispersed pulse.}
    \label{fig:MTP1_sum_profile}
\end{figure*}

The first MeerTRAP source was discovered on 17 September 2019 during commensal observations with the Thousand Pulsar Array project \citep[TPA,][]{2020johnston}. The source was discovered in one CB centered at RA~18:43:31 and Dec~$-$07:57:22 with a S/N\footnote{Unless otherwise specified, all quoted S/N, DM, and pulse width values are those reported by the \texttt{ASTROACCELERATE single pulse detection algorithm.}} of 12 and a DM of 255~pc~cm$^{-3}$. The known pulsars closest to the detection beam were PSR~J1843$-$0806 and PSR~J1842$-$0800, which have DMs of 215.8(9) and 188.6~pc~cm$^{-3}$ respectively and are each about 10\arcmin\ away from the beam centre \citep[][]{2004hobbs,2015ng}. Thus, after manual RFI cleaning and inspection of the candidate filterbank we concluded that the source was both new and astrophysical. The discovery pulse, dedispersed and with a RFI mask applied, is shown in Figure~\ref{fig:profiles}.

The discovery coordinates were re-observed during test observations by MeerKAT on 7 November 2019, during which time 32 further pulses from the source were detected over the course of 20 minutes. A similar observation on 23 March 2020 delivered a further 33 pulses in 20 minutes. For the follow-up observation the CBs were packed much more closely, overlapping at 98 per cent of their maximum sensitivity. As a result, most of the pulses were detected in many CBs at once, varying from one to 44 at a time, depending on the brightness of the pulse. The tight tiling of the beams was done with the aim of testing localising of sources. Using a custom software implementation, called \texttt{SeeKAT}\footnote{\url{https://github.com/BezuidenhoutMC/SeeKAT}} (Bezuidenhout et al.~in prep), of the localisation method first put forward in \citet{2015obrocka}, we were able to localise PSR~J1843$-$0757 to RA~18:43:33.01 and Dec~$-$07:57:36, with a 1-$\sigma$ statistical uncertainty of 1\arcsec\ in both RA~and Dec. MeerKAT has not pointed near the source since the discovery and the tests.

PSR~J1843$-$0757 has been observed periodically by the Lovell telescope since its discovery. This amounted to $\sim10$ hours of observations from September 2020 to June 2021. In that time, 51 pulses were positively identified. Single pulses were also identified 101 times in $\sim$60~minutes worth of Parkes observations of its coordinates. The disparity in detection rates between observatories is partially a reflection of the sensitivities of the respective instruments, but may also be affected by clustering of the pulses in time.

Figure~\ref{fig:MTP1_sum_profile} shows the dynamic spectrum and folded pulse profile summed over all detections of PSR~J1843$-$0757 using the MeerKAT (left), Lovell (middle), and Parkes (right) telescopes. Visible in all three profiles, but especially prominent in the MeerKAT detections, is the existence of a leading component, preceding the main pulse by $\sim$10 milliseconds.

The detection dates of all known PSR~J1843$-$0757 pulses using any of the telescopes are indicated by the markers in Figure\,\ref{fig:MTP1_dets}, and the vertical lines below show observations of the source with the respective telescopes. It is important to note that whereas the indicated Parkes and Lovell telescope observations were such that the boresight coordinates coincided with the best known source coordinates, the MeerKAT observations are those for which the source was coincidentally within the FWHM of the MeerKAT primary beam. The source was placed directly at MeerKAT boresight on only the two test occasions. Hence the detection rates of the source with the Lovell and Parkes telescopes should not be compared directly to that of MeerKAT. 

There is evidence of irregular clustering in the times-of-arrival (ToAs) of pulses from PSR~J1843$-$0757 during observations when the source was detected. To investigate possible clustering in the wait time between subsequent pulses we employed the method used to identify clustering in repeating pulses from FRB121102 in \citet{2018Oppermann}. If the detections of PSR~J1843$-$0757 single pulses were a Poisson-distributed random process, the wait times $\delta$ would correspondingly take on an exponential distribution, i.e.,
\begin{equation}\label{eq:exp}
\mathcal{P}(\delta \mid r)=r \mathrm{e}^{-\delta r},
\end{equation}
where $r$ is the constant pulsation rate. However, as \citet{2018Oppermann} describes, the wait time distribution can be generalised as a Weibull function of the form
\begin{equation}\label{eq:weibull}
\mathcal{W}(\delta \mid k, r)=k \delta^{-1}[\delta r \Gamma(1+1 / k)]^{k} \mathrm{e}^{-[\delta r \Gamma(1+1 / k)]^{k}},
\end{equation}
where
\begin{equation}
\Gamma(x) = \int_{0}^{\infty} t^{x-1} \mathrm{e}^{-t} \mathrm{d} t 
\end{equation}
is the gamma function. Eq.~\ref{eq:weibull} has a shape parameter $k$, and in the case that $k=1$, 
Eq.~\ref{eq:weibull} reduces back to the exponential function Eq.~\ref{eq:exp}. Hence, when the Weibull distribution is fit to a sample of pulse wait times, any deviation of the best-fit $k$ from unity indicates a departure from a Poissonian random process. In that case, $k<1$ specifies that a pulse is more likely to follow shortly after another pulse than would be the case for a Poissonian distribution, and the pulses can therefore be said to be clustered in time. Conversely, a value of $k>1$ would mean that subsequent pulses are less likely to follow in close succession than would be expected if they followed a Poissonian distribution with the same burst rate.

\begin{figure*}
    \centering
    \includegraphics[width=\textwidth]{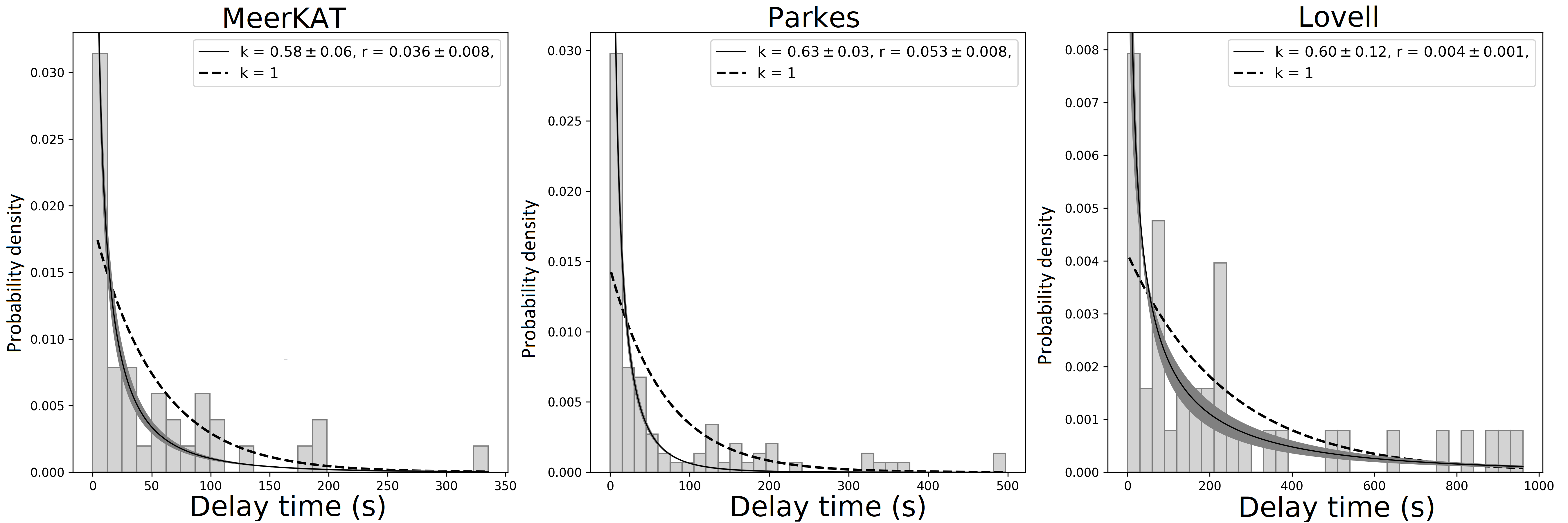}
    \caption{The probability density function of waiting times between subsequent detections of PSR~J1843$-$0757 using different telescopes. Weibull distributions with different shape parameters ($k$) are overlaid. In each panel the dashed line shows the best-fit Weibull distribution assuming $k=1$, corresponding to an exponential function. The solid lines (with 1-$\sigma$ errors shaded) indicate the best-fit Weibull distribution with $k$ and the burst rate ($r$) as free parameters. The derived values of $k$ and $r$ are noted in the legends.}
    \label{fig:MTP1_wait}
\end{figure*}

To fit the Weibull distribution to the observed waiting times we used the data analysis tool \texttt{scipy.optimize.curve\_fit}, which applies a non-linear least squares test using the Levenberg-Marquardt algorithm described in \citet[][]{1978more}. For each telescope the wait times were divided into 20 bins and the probability density function (PDF) of the distribution computed, to which Eq.~\ref{eq:weibull} was then fit. This produced optimal parameters (with 1-$\sigma$ errors also provided by the tool) of $k=0.58\pm0.06$ and $r=0.036\pm0.008$~s$^{-1}$for MeerKAT pulses, $k=0.63\pm0.03$ and $r=0.053\pm0.008$~s$^{-1}$for Parkes pulses, and $k = 0.63\pm0.03$ and $r=0.004\pm0.001$~s$^{-1}$for Lovell pulses. Figure~\ref{fig:MTP1_wait} shows the sample of PSR~J1843$-$0757 wait times fit to the Weibull distribution in Eq.~\ref{eq:weibull}.

As the dashed lines in Figure~\ref{fig:MTP1_wait} show, an exponential function (optimised for $r$) does not fit the wait times well for observations using any of the telescopes, suggesting that PSR~J1843$-$0757 pulses are clustered in time. It should be noted that while the shape parameter is expected to be consistent using different telescopes, the burst detection rate should be lower for less sensitive telescopes. From these fits to the Weibull distribution we thus conclude that PSR~J1843$-$0757's pulses are intrinsically clustered in time, such that they are grouped more tightly than would be expected for a Poissonian process.

To measure the arrival times of PSR~J1843$-$0757's pulses, we first used \texttt{dspsr} to produce folded pulse profiles for each data set by assuming an approximate period of 2.0~s derived from an arrival time differencing method\footnote{\url{https://github.com/evanocathain/Useful_RRAT_stuff}}. The folded data were then processed with the \texttt{PSRCHIVE} \citep[][]{2004Hotan} library for pulsar analysis. For each set of observations we created an analytic template profile from the cleaned pulse profile, averaged over all detections, using the \texttt{paas} tool. We then used \texttt{pat} to produce ToAs for all pulses. Using the ToAs from all three telescopes and the \texttt{TEMPO2}\footnote{\url{https://bitbucket.org/psrsoft/tempo2/}} pulsar timing software \citep{2006hobbs}, we established a long-term timing solution for the source. We measure a period of $P=2.03194008516(9)$~s, and a period derivative of $\Dot{P}=4.13(3)\times10^{-15}$. This implies a characteristic magnetic field strength of $B = 3.2\times10^{19}\sqrt{P\Dot{P}} \simeq 2.9\times10^{12}$~G and a characteristic age of $\tau = P/2\Dot{P} \simeq 7.8\times10^{6}$~years. The timing also provided a source position of RA 18:43:33.06(2) and Dec $-$07:57:33(2), which is within errors of what we derived using \texttt{SeeKAT}. This serves to reinforce confidence in our tied-array beam localisation method that was also used to localise PSR~J1152$-$6056 and PSR~J0930$-$1854.

\begin{figure*}
    \centering
    \includegraphics[width=0.98\textwidth]{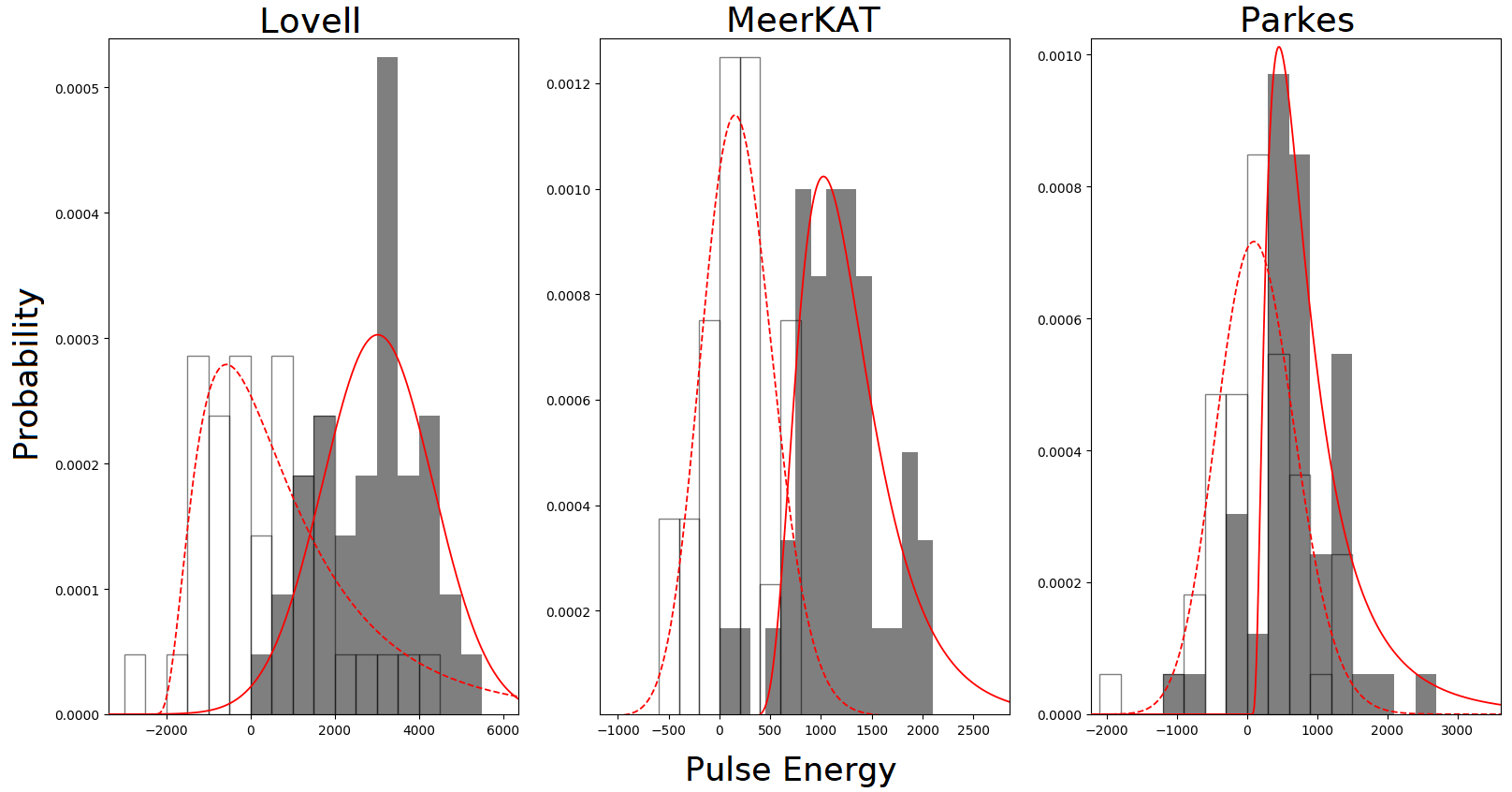}
    \caption{The distribution of pulse energies recorded for single pulse detections of PSR~J1843$-$0757 made using the Lovell (left), MeerKAT (middle), and Parkes (right) telescopes. The pulse energies are in arbitrary units. In each case the solid bars indicate energy counts of the on-pulse region of the pulse profile, while the outlined bars indicate the same for the off-pulse region. The red lines are the best-fit log-normal fit to the distributions.}
    \label{fig:MTP1_PED}
\end{figure*}

Figure~\ref{fig:MTP1_PED} shows the distribution of pulse energies of all PSR~J1843$-$0757 single pulse detections using the MeerKAT, Lovell, or Parkes telescopes. The solid bars indicate the on-pulse distribution, and the outlines represent off-pulse distributions. These were derived using the \texttt{PSRSALSA}\footnote{\url{https://github.com/weltevrede/psrsalsa}} software package \citep{2016weltevrede}. The offset of the average on-pulse energy from the average off-pulse energy clearly indicates the presence of single pulses. The figure also illustrates, especially in the case of the Parkes observations, the difficulty in distinguishing detections from non-detections, as there's a significant overlap in energy. The red lines show the best-fit log-normal function for each pulse energy distribution determined using \texttt{scipy.optimize.curve\_fit}. The $\chi^{2}$ test statistic for each fit calculated using the \texttt{scipy.stats.chisquare} \texttt{python package} is listed in Table~\ref{table:MTP1_PED_fits}.

\begin{table}
\centering
\caption{The $\chi^{2}$ values for the log-normal fit to each of the pulse energy distributions shown in Figure~\ref{fig:MTP1_PED}.}\label{table:MTP1_PED_fits}
\begin{tabular}{lrr}
\toprule
Telescope & Off-pulse $\chi^{2}$ & On-pulse $\chi^{2}$\\
\midrule
Lovell & 5.5 & 16.1\\
MeerKAT & 7.8 & 2.0\\
Parkes & 64.7 & 49.3\\
\bottomrule
\end{tabular}
\end{table}
\vspace{-1em}

\subsection{MTP0002/PSR~J1152$-$6056}
\label{subsec:MTP2}

\begin{figure}
    \centering
    \includegraphics[width=0.48\textwidth]{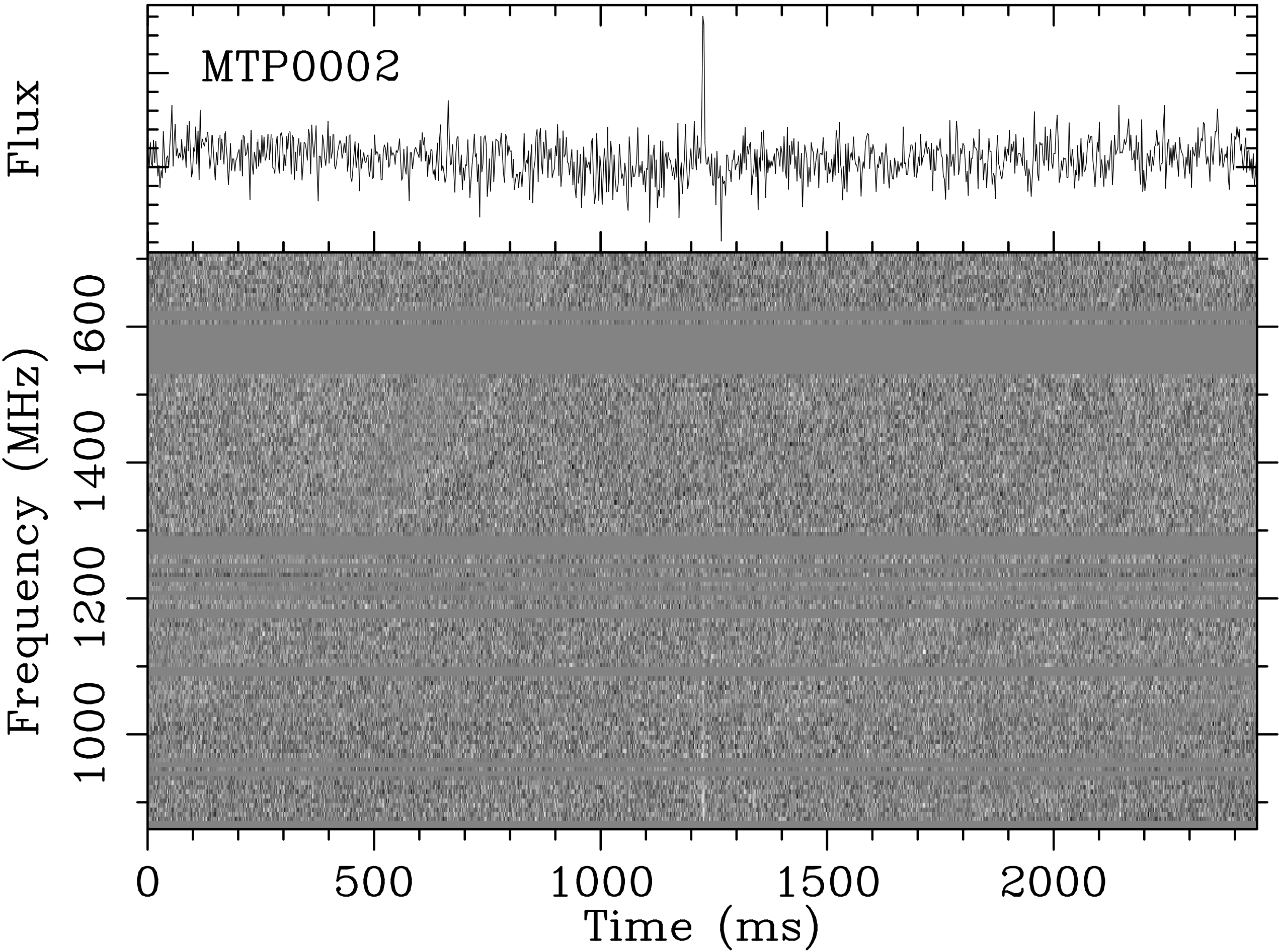}
    \caption{Dynamic spectrum (bottom) and pulse profile (top) of PSR~J1152$-$6056 averaged over 9 MeerKAT detections of the source.}
    \label{fig:MTP2_sum_profile}
\end{figure}

PSR~J1152$-$6056 was discovered on 15 March 2020 in a CB centred at RA~11:52:26.93 and Dec $-$60:56:38.4. The candidate pulse, with a DM of 381~pc~cm$^{-3}$, a S/N of 9, and a width of 6.12~ms, was visually confirmed and the filterbank file saved. During an observation two days later, when the TUSE CBs were tiled with 84 per cent overlap on the most likely position, six more pulses were seen. In the middle of the series of pulses the tiling pattern was refreshed so that the CBs changed position slightly. Consequently, while the first three pulses were detected in a CB at RA~11:52:37.27 and Dec~$-$60:56:22, the last three were seen at RA~11:52:37.19 and Dec~$-$60:56:24.7. Finally, in almost two hours of MeerKAT observations nearby to PSR~J1152$-$6056, there have been two additional detections, both within three minutes of each other on 6 July 2021 in a CB centred on RA~11:52:36.89 and Dec~$-$60:56:37.6.

Using the nine pulses from the source we have localised it to a best-fit position of RA~11:52:37.44 and Dec~$-$60:56:22.60 with 1-$\sigma$ errors of 2\arcsec\ in RA~and 1\arcsec\ in Dec. There are no known sources with a similar DM within 2$^{\circ}$ of these coordinates recorded in the pulsar or RRAT catalogs.

PSR~J1152$-$6056 has also been periodically re-observed using the Parkes telescope. No further pulses were discovered in all 2 hours of Parkes UWL observations. As a result, the seven MeerKAT detections are the only known pulses from the source. The source is below the declination limit of the Lovell telescope.

Using the same arrival time differencing method used to derive an approximate period for PSR~J1843$-$0757 using the arrival times of the six pulses detected on 17 March 2020 (average epoch 58925.7279484) revealed a tentative periodicity of around 2.45~s. Using \texttt{TEMPO2} we were able to confirm a period of 2.449000(6)~s. However, because of the lack of detections beyond these intial ones, we were not able to determine a period derivative. Figure~\ref{fig:MTP2_sum_profile} shows the profile of PSR~J1152$-$6056 averaged over the nine detections.

\subsection{MTP0003}

MTP0003 was only detected once with MeerKAT, on 12 March 2020 in one CB centred on the coordinates RA~17:42:35 and Dec~$-$14:01:39 with a S/N of 8 and DM of 124~pc~cm$^{-3}$. 
The nearest transient source with a similar DM is PSR~J1743$-$1351, with a DM of 116.30(3)~pc~cm$^{-3}$ and flux density at 1.4~GHz of 0.5(1)~mJy, located at RA~17:43:37.622 and Dec~$-$13:51:38.0, which is 0.303$^{\circ}$ away from the MTP0003 detection beam centre \citep[][]{2004hobbs}. Given the DM disparity, the distance from PSR~J1743$-$1351's known coordinates, and the lack of detections in surrounding CBs, we consider MTP0003 to be most likely a distinct source, although we cannot rule out the possibility of it being a sidelobe detection of PSR~J1743$-$1351.

The source has not been observed again by MeerKAT or the Lovell telescope at the time of writing. Thirty minutes of observations with the Parkes UWL receiver have not resulted in any further detections. 

\subsection{MTP0004}

This source was seen only once, in a CB centred on RA~18:24:52 and Dec~$-$13:31:19 on 12 March 2020. The pulse had a DM of 212~pc~cm$^{-3}$ and was detected with a S/N value of 14. The nearest source with a similar DM is PSR~J1826$-$1334, located at RA~18:26:13.175(3) and Dec~$-$13:34:46.8(1) (i.e.~0.334$^{\circ}$ away from the aforementioned beam), which has a DM of 231.0(10)~pc~cm$^{-3}$ and a flux density at 1.4~GHz of 4.7(2)~mJy \citep[][]{1986Clifton}. Due to the DM difference and the distance of PSR~J1826$-$1334 from the MTP0004 detection beam, MTP0004 is unlikely to have been a pulse from PSR~J1826$-$1334.

We have not identified any further pulses in 44 minutes of MeerKAT observations, one hour of Lovell observations, and 30 minutes of Parkes UWL observations.

\subsection{MTP0005/PSR~J1840$-$0840}

On 23 March 2020 two successive pulses were detected in the IB centred at RA~18:42:55 and Dec~$-$08:00:54, 5.33~s apart. The pulses had S/N values of 10 and 12, respectively, and both had a DM of 299~pc~cm$^{-3}$. A third pulse was detected on 31 October 2021, this time during a MeerKAT observation with boresight coordinates of RA~18:40:33 and Dec~-08:09:03. This pulse was also seen in the IB, and had a DM of 299~pc~cm$^{-3}$ and S/N of 9.

Although this source was at first considered a newly discovered transient, the IB detections now seem likely to have been of PSR~J1840$-$0840, despite the fact that the pulsar was 50\arcmin~from boresight at the time of the first detection and 32\arcmin~away at the time of the second. PSR~J1840$-$0840 was first discovered in the PMPS with a DM of 272(19)~pc~cm$^{-3}$ \citep[][]{2006lorimer}, and this DM value was used in all subsequent literature on the source that we have found. However, a check of timing data with the Lovell telescope (also at the discovery frequency of 1.4~GHz) gives a DM of 300(12)~pc~cm$^{-3}$, and a reanalysis of a Parkes telescope observation gives a DM of 300(8)~pc~cm$^{-3}$ (S.\ Johnston priv. comm.). The DM error is large because of the relatively wide profile of this slow pulsar; \citet[][]{2006lorimer} gives a W$_{\mathrm{50}}$ value of 188~ms. 

PSR~J1840$-$0840 has a period of 5.3093766847(20)~s \citep[][]{2006lorimer}, which differs from the time between subsequent MTP0005 pulses by about 21~ms. Combined with the wide profile of the pulsar, this disparity can be explained by the source being highly variable in flux and exhibiting drifting subpulses \citep[][]{2017gajjar}. 

No further pulses were seen in 146 minutes of MeerKAT observations or 20 minutes of Parkes UWL observations. All of the Lovell observations of the source were heavily affected by RFI. The follow up observations were all pointed at the boresight coordinates of the first MeerTRAP IB detection, since the identification with PSR~J1840$-$0840 had not been made at the time of observing.

\subsection{MTP0006}
One pulse of MTP0006 was seen on 12 June 2020. The pulse was detected in three CBs simultaneously, although they were far apart. Detections were made in beams that were centred at RA~18:40:34 and Dec~$-$09:12:44, RA~18:39:51 and Dec.~$-$09:10:00, and RA~18:40:57 and Dec~$-$08:48:31, respectively. The pulses all had S/N values of 8, just above the detection limit. The DMs of the detected pulses also varied slightly, being 302, 304, and 295~pc~cm$^{-3}$, respectively. The wide spatial dispersion of the detection beams most likely indicates that the detections were made in distinct side-lobes of the CBs' point spread functions (PSFs), and that the true position of the source is somewhere outside of the CB tiling region. We have therefore not been able to localise the source any more precisely.

We could identify only two known sources in the vicinity of these beams with a similar DM. PSR~J1832$-$0827, which is 1.92$^{\circ}$ away from the nearest of the detection beams, has a DM of 300.869(10)~pc~cm$^{-3}$ and 1.4~GHz flux density of 4(3)~mJy \citep[][]{2019parthasarathy}. PSR~J1834-0731 is 2.1$^{\circ}$ away from the nearest of those beams, and has a DM of 294.0(9)~pc~cm$^{-3}$ and a flux density of 1.3(1)~mJy \citep{2019parthasarathy}. The distance of these sources from the beams in which MTP0006 were detected, combined with their relatively low flux densities, makes it improbable for MTP0006 to have been detections of a pulse from one of these sources.

No further detections of MTP0006 have been made in 23 minutes of MeerKAT observations nearby any of the three CBs' coordinates, 3.5 hours of Lovell observations, and 25 minutes of Parkes UWL observations. For the Lovell and Parkes observation the telescopes were pointed at the coordinates of the CB detection with the highest S/N.

\subsection{MTP0007/PSR J0930$-$1854}

\begin{figure*}
    \centering
    \includegraphics[width=0.9\textwidth]{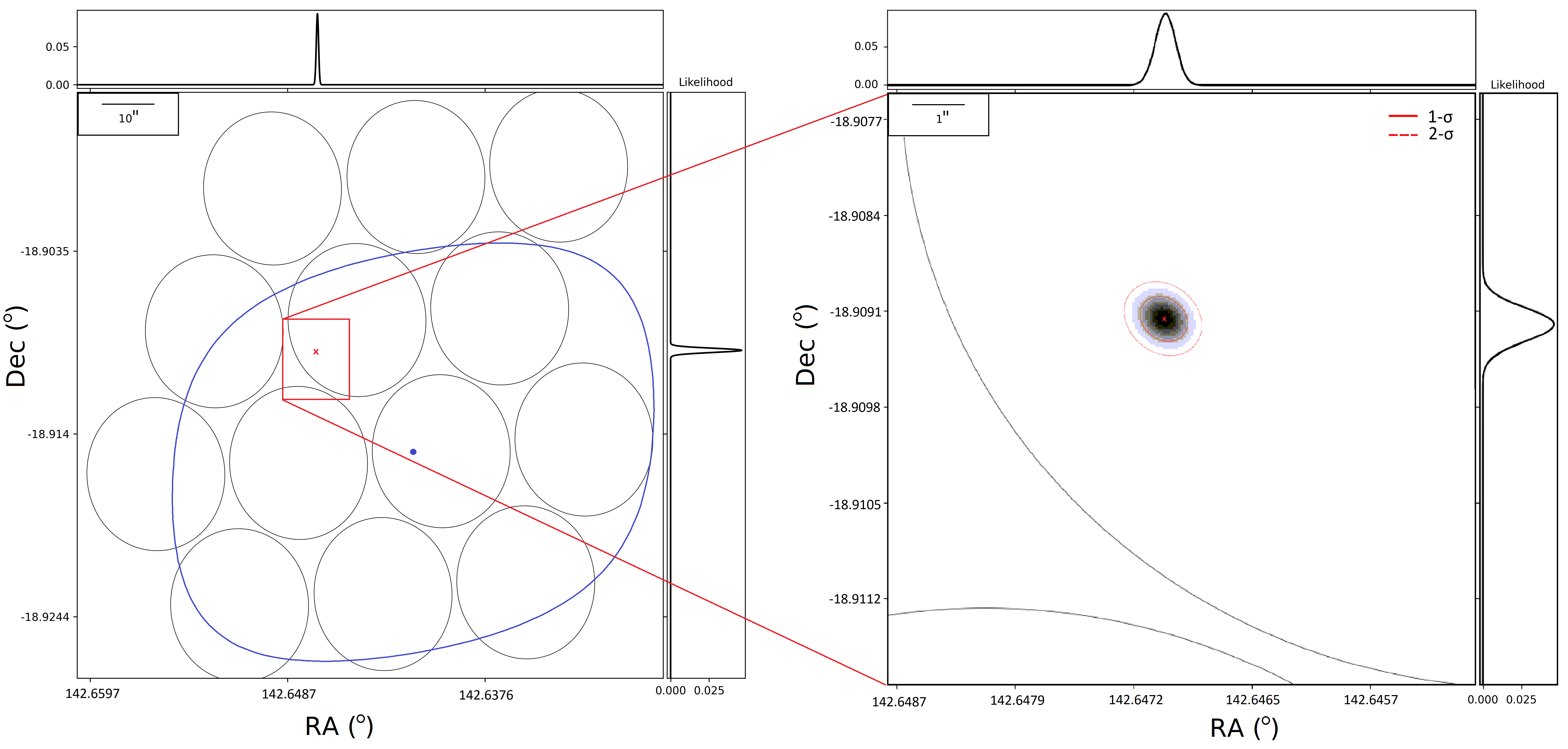}
    \caption{Tied-Array Beam Localisation of PSR~J0930$-$1854 based on the source's detection in thirteen MeerKAT CBs simultaneously. The large, blue contour shows the 25\% level of the CB in which the source was discovered on 18 June 2020. The smaller black contours are centred on the position of each of the thirteen beams the source was detected in during a follow up observation on 24 May 2021. In the test observations the beams were arranged to overlap at 84\% of their maximum sensitivity. The right panel is an enlarged version of the contents of the red square in the left panel. The red cross shows the most likely position of PSR~J0930$-$1854 based on the distribution of S/N values of the detections in respective beams, viz.~RA~09:30:35.27 and Dec~$-$18:54:33.1. The red contours show 1-$\sigma$ (solid) and 2-$\sigma$ (dashed) errors on this position. The scale of each of panel is indicated in the legends.}
    \label{fig:MTP7_localisation}
\end{figure*}

This source was first detected on 18 June 2020. A single pulse was seen in a CB centred at RA~09:30:34 and Dec~$-$18:54:54 with a DM of 33~pc~cm$^{-3}$ and a S/N of 9. In 250 minutes of reobservations with MeerKAT one more pulse from the source was detected, on 24 May 2021 in 11 adjacent CBs as well as the IB simultaneously. Based on these detections we were able to localise this source using \texttt{SeeKAT} to RA~09:30:35.27(2) and Dec~$-$18:54:33.1(8), where the one-$\sigma$ errors are given in the parentheses. The result of this localisation is shown in Figure~\ref{fig:MTP7_localisation}. Unfortunately, 3.5 hours with the Lovell telescope, and one hour with the Parkes UWL have not resulted in any other detections.

\subsection{MTP0008/PSR~J1901+0254}

\begin{figure*}
    \centering
    \includegraphics[width=\textwidth]{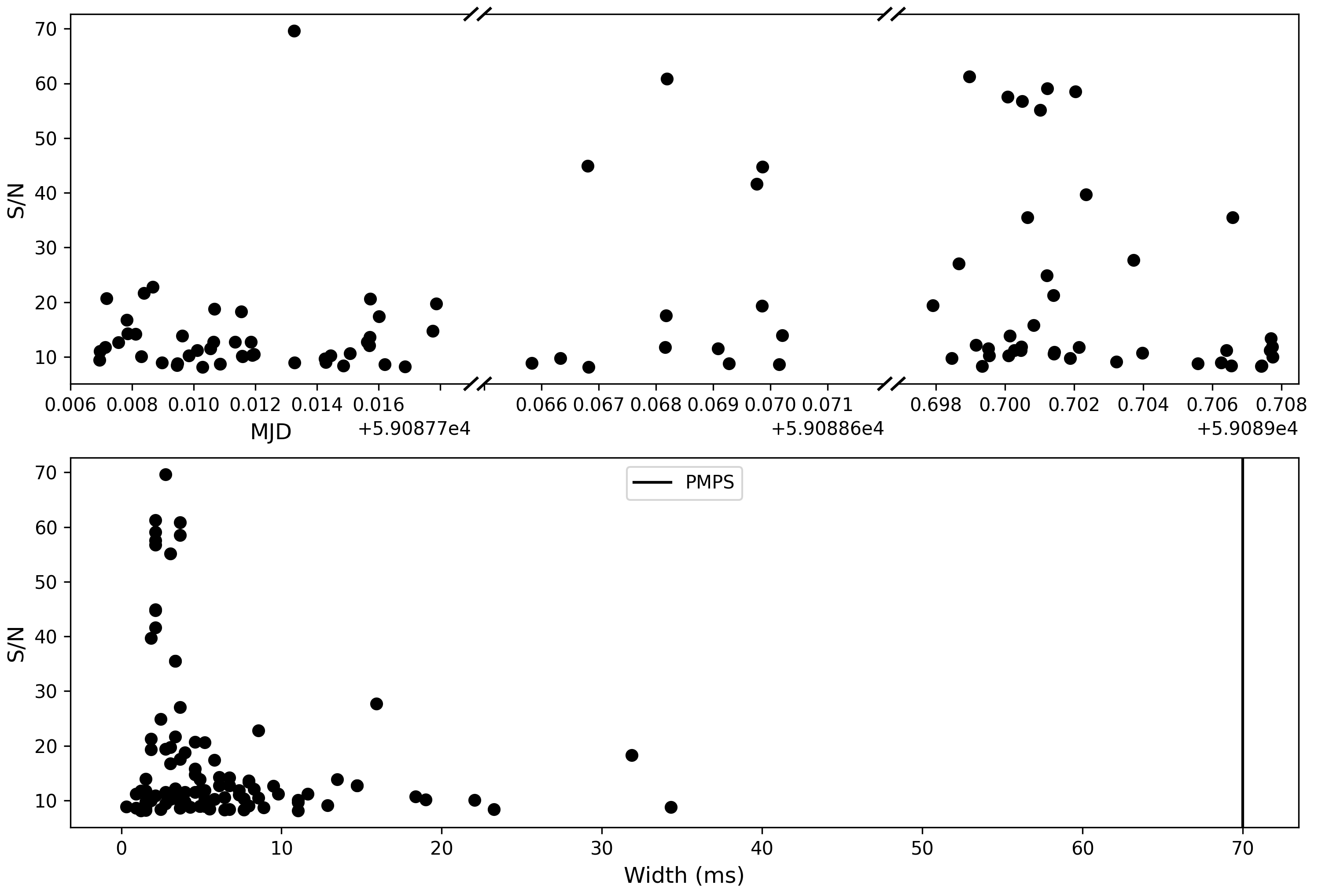}
    \caption{Arrival times (top) of pulses from PSR~J1901+0254 detected by MeerTRAP over $\sim$45 minutes. The breaks in the horizontal axis correspond to the three consecutive days that the observations were spread out over. S/N values vs.~pulse widths for the same pulses (bottom).}
    \label{fig:MTP8_Width}
\end{figure*}

\begin{figure*}
    \centering
    \includegraphics[width=\textwidth]{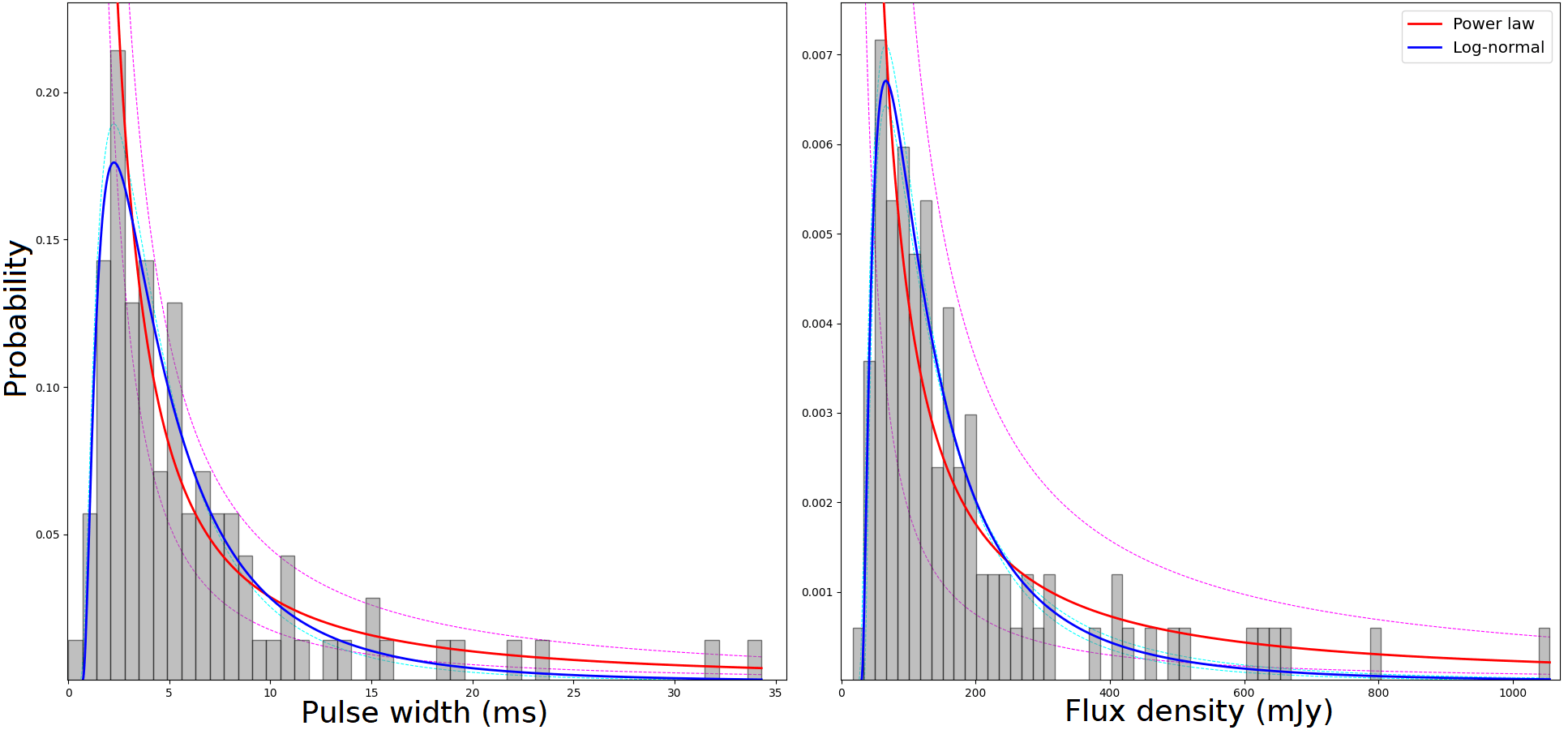}
    \caption{Pulse width distribution (left) and pulse energy distribution (right) of 100 PSR~J1901+0254 pulses detected by MeerTRAP. The flux densities are estimates derived using Eq.~\ref{eq:radiometer}. The blue and red lines show the best-fit log-normal and power law functions, respectively, with the dashed lines in each case indicating 1-$\sigma$ errors on the fit.}
    \label{fig:MTP8_PED}
\end{figure*}

\begin{table}
\centering
\caption{The $\chi^{2}$ values for the log-normal and power law functions fit to each of the pulse width and energy distributions shown in Figure~\ref{fig:MTP8_PED}.}\label{table:MTP8_PED_fits}
\begin{tabular}{lrr}
\toprule
Parameter & Log-normal $\chi^{2}$ & Power law $\chi^{2}$\\
\midrule
Width & 118.9 & 36.1\\
Energy & 108.9 & 28.1\\
\bottomrule
\end{tabular}
\end{table}

MTP0008 is another source that was only detected in the MeerKAT IB. It was also found on 18 June 2018, during an observation centred on RA~19:03:06 and Dec~+03:27:19. The pulse had a S/N of 8 and a DM of 181~pc~cm$^{-3}$. 

Initially it was thought that MTP0008 was a newly discovered source. The centre of the IB at the time of detection was located $\sim$41$\arcmin$ away from the nearest source with a similar DM, viz.~PSR~J1901+0254, which was first reported in the PMPS as having a DM of 185$\pm$5~pc~cm$^{-3}$ and a flux density at 1400~MHz of 0.58~mJy \citep[][]{2004hobbs}. Because of the source's distance from the IB centre, the DM difference, and the relatively low flux density of the source, we originally considered it unlikely that MTP0008 was a pulse from PSR~J1901+0254.   

However, during MeerKAT observations of PSR~J1903+0327 \citep[DM 297.5~pc~cm$^{-3}$,][]{2008champion} on 26--28~August 2020 ammounting to $\sim$45 minutes, we detected 106 pulses in the IB and CBs from PSR~J1901+0254. The brightest pulses were in CBs a few arcseconds away from the pulsar's catalog position, RA~19:01:15.67 and Dec~+02:54:41, and had DMs from 179 to 182~pc~cm$^{-3}$. A simple timing analysis also established that the pulses' ToAs conform to PSR~J1901+0254's 1.2996934495(3)~s spin period. This leads us to believe that MTP0008 was likely an IB detection of PSR~J1901+0254 outside of the reach of the CBs.

Interestingly, the pulses were significantly brighter than would be expected given the recorded flux for PSR~J1901+0254, with S/N values up to 61. They were also extremely variable, with pulses in the nearest beam to the source position registering S/N values\footnote{Unfortunately we did not save any filterbank data for these pulses, so we could not obtain any more precise DM, pulse width, or S/N measurements than what \texttt{ASTROACCELERATE} provides in our stored candidate plots.} ranging in S/N from 8 to 61. In order to correct the S/N values for the source's position within the CB, we simulated the CB PSFs using the beam synthesis simulation code \texttt{MOSAIC} \citep[][]{2021chen} and estimated the fractional sensitivity of the beam at the position of PSR~J1901+0254; we then divided the S/N values by this fraction. We also applied a small correction for the CBs' position within the MeerKAT primary beam.

To calculate approximate peak pulse flux densities from the corrected S/N values we used the radiometer equation \citep[][]{1985Dewey}
\begin{equation}\label{eq:radiometer}
    S \simeq\frac{\text { S/N } T_{\text {sys }}}{G \sqrt{n_{\mathrm{p}} W \Delta f}},
\end{equation}
where $G$ is the telescope's gain, $n_\mathrm{p}$ is the number of polarisation channels recorded, $W$ is the width of the pulse (de-dispersed and compensating for any other widening effects), and $\Delta f$ is the bandwidth. Instrumental parameters for MeerKAT from \citet{2020bailes} are $n_p = 2$, $\Delta f = 856$~MHz, and $G = 1.575$~K/Jy (modified from 2.8~K/Jy since only 36 out of 64 MeerKAT dishes were used for the observations). The telescope's system temperature $T_{\mathrm{sys}}$ is given by
    $T_{\mathrm{sys}}$ =  $T_{\mathrm{rec}} +  T_{\mathrm{sky}}$,
with $T_{\mathrm{rec}} = 18$~K the receiver temperature for L-band observations \citep[e.g.~][]{2021ridolfi}, and the sky temperature at the source coordinates at the centre of the L-band $ T_{\mathrm{sky}} \sim 7.8$~K determined using a \texttt{python} implementation\footnote{\url{https://github.com/jeffzhen/gsm2016}} of the diffuse galactic radio emission model of \citet[][]{2017zheng}.

Inserting these parameters as well as the S/N and pulse width for all CB detections into Eq.~\ref{eq:radiometer}, we obtain a very wide range of flux density estimates, from $\sim$10$^{-2}$ to 1~Jy, with the mean value being 184.5~mJy. Compared to the 0.58(7)~mJy mean flux density found for PSR~J1901+0254 in the PMPS \citep[][]{2004hobbs} and the 0.911~mJy found in the PALFA pulsar survey with the Arecibo telescope \citep[][]{2015lazarus}, the pulses detected by MeerTRAP are one to two orders of magnitude greater in flux density. One contributing factor is likely the width of the pulses, which were measured as ranging between about 0.3 and 34~ms (with the median being 4.4~ms), much smaller than the 70~ms W$_{50}$ width reported from the PMPS survey; \citet[][]{2015lazarus} did not cite a width figure. This may indicate the presence of brighter, narrower subpulses in the PSR~J1901+0254 pulse profile.

A second factor giving rise to the higher average flux density is the significant modulation in intrinsic pulse intensity, with some pulses having a flux density an order of magnitude higher than the median value. Figure~\ref{fig:MTP8_PED} shows the distribution of pulse widths (left) and beam corrected flux densities (right) detected by MeerTRAP from PSR~J1901+0254. The blue and red lines show the best-fit log-normal and power law distributions, respectively, as determined using \texttt{scipy.optimize.curve\_fit}. Table~\ref{table:MTP8_PED_fits} indicates the corresponding $\chi^{2}$ values calculated using \texttt{scipy.stats.chisquare}. Note that to account for possible incompleteness of the observed distributions on the lower end of the pulse width and flux density measurements, the power law fit was only performed for widths greater than 3~ms and flux densities greater than 50~mJy (i.e.~for values past the peak of the distribution). In both cases the log-normal function provides a closer fit to the observed distribution than the power law function does.

Besides the MeerTRAP pulses all being considerably narrower and brighter than the average PMPS and PALFA pulses, a few extremely bright pulses can be seen in the tail of the distribution in Figure~\ref{fig:MTP8_PED}. A 2009 search for single pulses from known pulsars with the Parkes Multibeam receiver did not report any single pulses from PSR~J1901+0254 \citep[][]{2010burke-spolaor}. The non-detection in this survey of the very bright PSR~J1901+0254 pulses seen using MeerKAT suggests a possible evolution in the pulse energy distribution of the source, although a single pulse study is required to establish this.

The brightest pulses may be similar to giant pulses from pulsars such as the Crab pulsar \citep[][]{1995lundgren}. Given PSR~J1901+0254's period of $P=1.2996934495$~s and period derivative $\Dot{P} = 4.6\times10^{-16}$ from \citet[][]{2004hobbs}, its characteristic age is $\tau = P/2\Dot{P} \simeq 4.5\times10^{7}$~years. Its magnetic field at the light cylinder, where corotation would occur at the speed of light, is $B_{lc} \sim 3\times10^{8}~P^{-5/2}\Dot{P}^{1/2} = 3.3$~G. These properties are not atypical for a canonical pulsar, unlike the Crab pulsar, which is extremely young ($\tau=1.2\times10^{3}$~years) and has an uncommonly strong magnetic field at the light cylinder of $B_{lc} \sim 10^6$~G. The other pulsars that have been found to emit giant pulses, including PSR~B1937+21 \citep[][]{1996cognard}, PSR~B1821$-$24 \citep[][]{2001romani}, and PSR~B0450$-$69 \citep[][]{2003johnston}, all have markedly different periods and period derivatives, but have in common an unusually strong magnetic field at the light cylinder. This paramater has thus often been used as an indicator of giant pulse emissivity \citep[e.g.~][]{1996cognard}. Since PSR~J1901+0254 has a relatively weak light cylinder magnetic field, it is unlikely that its unusually bright pulses represent a similar phenomenon to Crab giant pulses.

Aside from observations of PSR~J1901+0254 during the PMPS and PALFA campaigns, the source was also a target of a survey carried out on radio pulsars in supernova remnants (SNRs) using the XMM-Newton X-ray space telescope \citep[][]{2014bogdanov}. PSR~J1901+0254 was targeted as a relatively old, nearby pulsar coincident with the SNR~G36.6$-$0.7 \citep{1987fuerst}, to investigate the possibility of it being a central compact object (CCO), which are expected to have anomalously high thermal X-ray luminosities and characteristic ages older than the coincident SNRs. However, no X-ray emission was detected from PSR~J1901+0254 in 16.9~ks exposure with XMM-Newton, implying a bolometric X-ray luminosity of $L_x \lesssim 10^{32}~\textrm{ergs}~\textrm{s}^{-1}$. This suggests that PSR~J1901+0254 is likely a regular pulsar unrelated to the SNR.

\subsection{MTP0009}
MTP0009 was discovered on 19 April 2020 in a MeerKAT CB located at RA~06:27:15.10 and Dec~06:56:26.7. The pulse was just above the detection threshold, with a S/N of 8, and a DM of 262~pc~cm$^{-3}$. No further pulses have been seen in 1.5 hours of MeerKAT, 2 hours of Lovell, or 1.5 hours of Parkes UWL observations.

It is notable that the Galactic DM contribution along MTP0009's line of sight differs significantly between the \texttt{NE2001} and \texttt{YMW16} electron density models; the former predicts a maximum contribution of 184~pc~cm$^{-3}$, while the latter predicts a maximum of 304~pc~cm$^{-3}$. The DM distances are $>$42.1~kpc and 7.0~kpc assuming the \texttt{NE2001} and \texttt{YMW16} models, respectively. There are no known transients with a similar or higher DM within ten degrees of MTP0009.

The nearby PSR~J0631+0646 has a DM of 192.2(2), which places it at a DM distance of $>42.2$~kpc using the \texttt{NE2001} model and 4.6 using the \texttt{YMW16} model \citep[][]{2018wu}. However, as \citet[][]{2018wu} notes, both radio and $\gamma$-ray pulses from PSR~J0631+0646 have been detected, and assuming the \texttt{NE2001} model DM contribution, the $\gamma$-ray conversion efficiency would exceed 100 per cent at a distance of about 6.7~kpc. \citet[][]{2018wu} therefore concludes that the \texttt{NE2001} model probably underestimates the free electron density along the line of sight of PSR~J0631+0646. Given that PSR~J0631+0646 is only 1.16 degrees away, the \texttt{YMW16} DM distance for MTP0009 is also expected to be more accurate, in which case MTP0009 is most likely Galactic. However, since the the measured DM is close to the maximum \texttt{YMW16} Galactic DM contribution along that line of sight, MTP0009 may yet exist in the Galactic halo or may be extragalactic. More observations of nearby sources are required to determine if the line of sight electron content can be more accurately ascertained. The detection of repeat pulses from MTP0009 may also be used to better 
localise the source and potentially determine an independent distance measurement by association with a host galaxy.

\subsection{MTP0010/PSR~J0723$-$2050}
A series of four pulses were detected in a MeerKAT CB at RA~07:23:02.74 and Dec~$-$20:50:04.8, all with a DM of 130~pc~cm$^{-3}$, within six minutes on 4 July 2020. Another three pulses were seen on 5 May 2021 at the same coordinates and with the same DM spread out over 12 minutes. Using the barycentred ToAs we were able to derive a pulse period of 711.54~ms with the \texttt{rratsolve} periodicity solver\footnote{\url{https://github.com/v-morello/rratsolve}}. However, because of the long gap without detections we were not able to phase connect the ToAs to find a timing solution with a period derivative.

\begin{figure}
    \centering
    \includegraphics[width=0.48\textwidth]{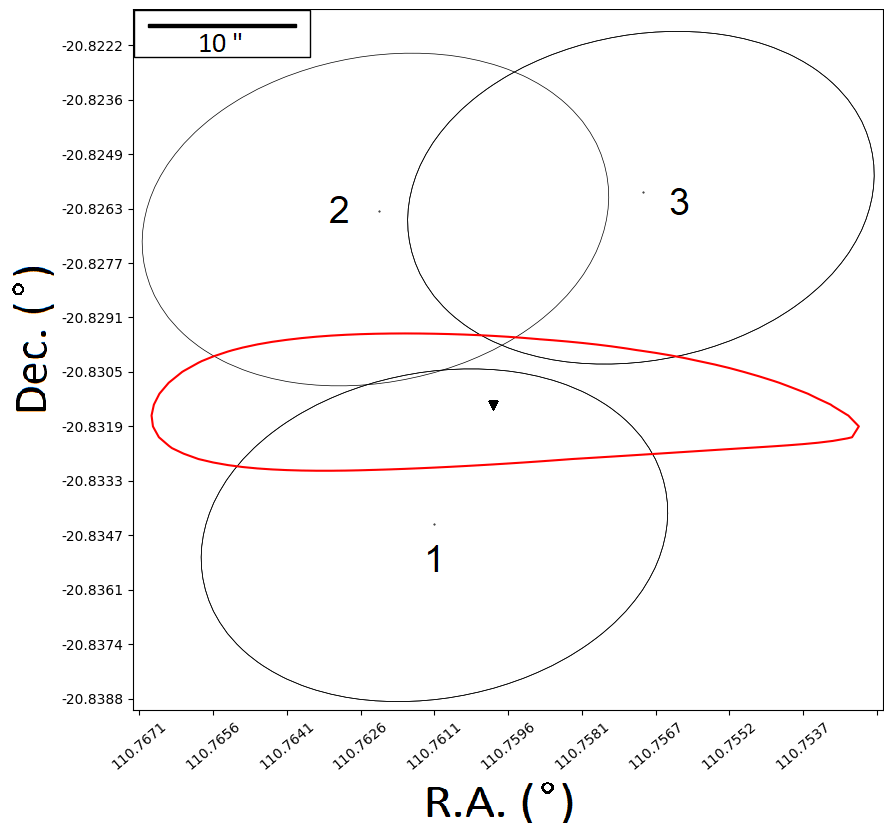}
    \caption{Tied-Array Beam Localisation of PSR~J0723$-$2050 based on the two pulses detected in three beams. The first pulse was seen in beams \textit{1} and \textit{2} simultaneously, while the second was seen in beams \textit{1} and \textit{3} simultaneously. The triangle indicates the best-fit position of RA~07:23:02.4 and Dec~$-$20:49:52(14). The red contour shows the 1-$\sigma$ uncertainty region which spans approximately 53\arcsec in RA~and 14\arcsec in Dec.}
    \label{fig:MTP10_localisation}
\end{figure}

Two of the three pulses on 4 July 2020 were detected in two CBs at once. Both pulses were detected in the CB located at RA~07:23:02.74 and Dec~$-$20:50:04.8, labelled \textit{1} in Figure~\ref{fig:MTP10_localisation}. The first pulse was also seen in the CB labelled \textit{2}, at RA~07:23:03.01 and Dec~$-$20:49:36.0, while the second pulse, which came 12.6~minutes later, was seen in both beam \textit{1} and \textit{3} located at RA~07:23:01.73 and Dec~$-$20:49:34.3. We were able to combine the CB localisation for the two pulses to derive a best-fit position of RA~07:23:02.4(18) and Dec~$-$20:49:52(7). These coordinates do not match the coordinates of any known transient with a comparable DM. The triangle in Figure~\ref{fig:MTP10_localisation} shows the best-fit coordinates, and the red contour is the 1-$\sigma$ uncertainty region. 

No further single pulses of the source were confirmed in the approximately one hour total MeerKAT observing time and one hour of Parkes UWL observations.

\subsection{MTP0011/PSR~J0943$-$5305}
PSR~J0943$-$5305 is the first MeerTRAP source to have been discovered using the MeerKAT UHF-band receiver. 

\begin{figure}
    \centering
    \includegraphics[width=0.48\textwidth]{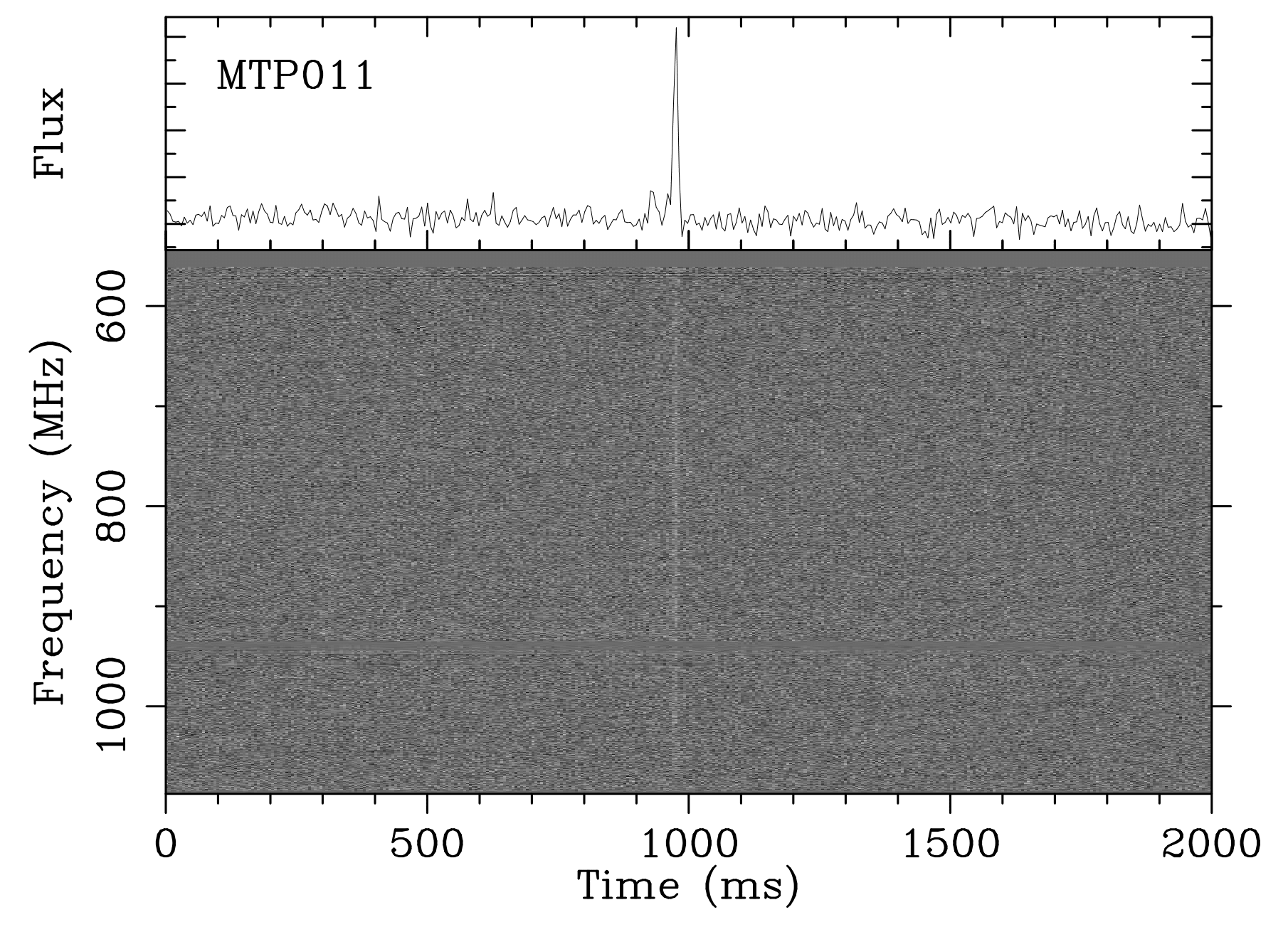}
    \caption{Intensity as a function of frequency and time (bottom) and frequency averaged profile (top) of PSR~J0943$-$5305 averaged over 7 MeerKAT detections of the source.}
    \label{fig:MTP11_sum_profile}
\end{figure}

Seven single pulses from this source were detected in a MeerKAT CB located at RA~09:43:08:09 and Dec~$-$53:05:17.7 in the span of about three minutes. The pulses, ranging in S/N from 9 to 30, all had a DM of 174~pc~cm$^{-3}$. Figure~\ref{fig:MTP11_sum_profile} shows the source's profile averaged over the seven detections. We were able to identify a likely periodicity of $\sim$1.734~s, although we have not been able to establish a complete timing solution.

No subsequent detections have been made in about 100 minutes worth of observations nearby to the source with MeerKAT and 85 minutes with the Parkes telescope. 

\subsection{MTP0012}

MTP0012 was also discovered with the MeerKAT UHF-band receiver. Only one pulse from the source has been observed during the approximately 25 minutes of exposure with MeerKAT. It was detected with a DM of 161~pc~cm$^{-3}$ and a S/N value of 21 in a CB at coordinates RA~19:14:43.43 and Dec~02:18:34.3. These coordinates are 1.4 degrees away from the nearest known compact object, PSR~J1909+0254, which has a DM of 171.734(9)~pc~cm$^{-3}$ \citep{2004hobbs}. Taking into account the brightness of the detected pulse, the distance from the known coordinates, the DM difference, and the broad frequency range that the pulse was visible over, we consider it unlikely for MTP0012 to have been a far sidelobe detection of PSR~J1909+0254.

Unfortunately we have not been able to obtain Lovell or Parkes observing time for MTP0012, so no further information on it is available.

\begin{figure*}
    \centering
    \includegraphics[width=\textwidth]{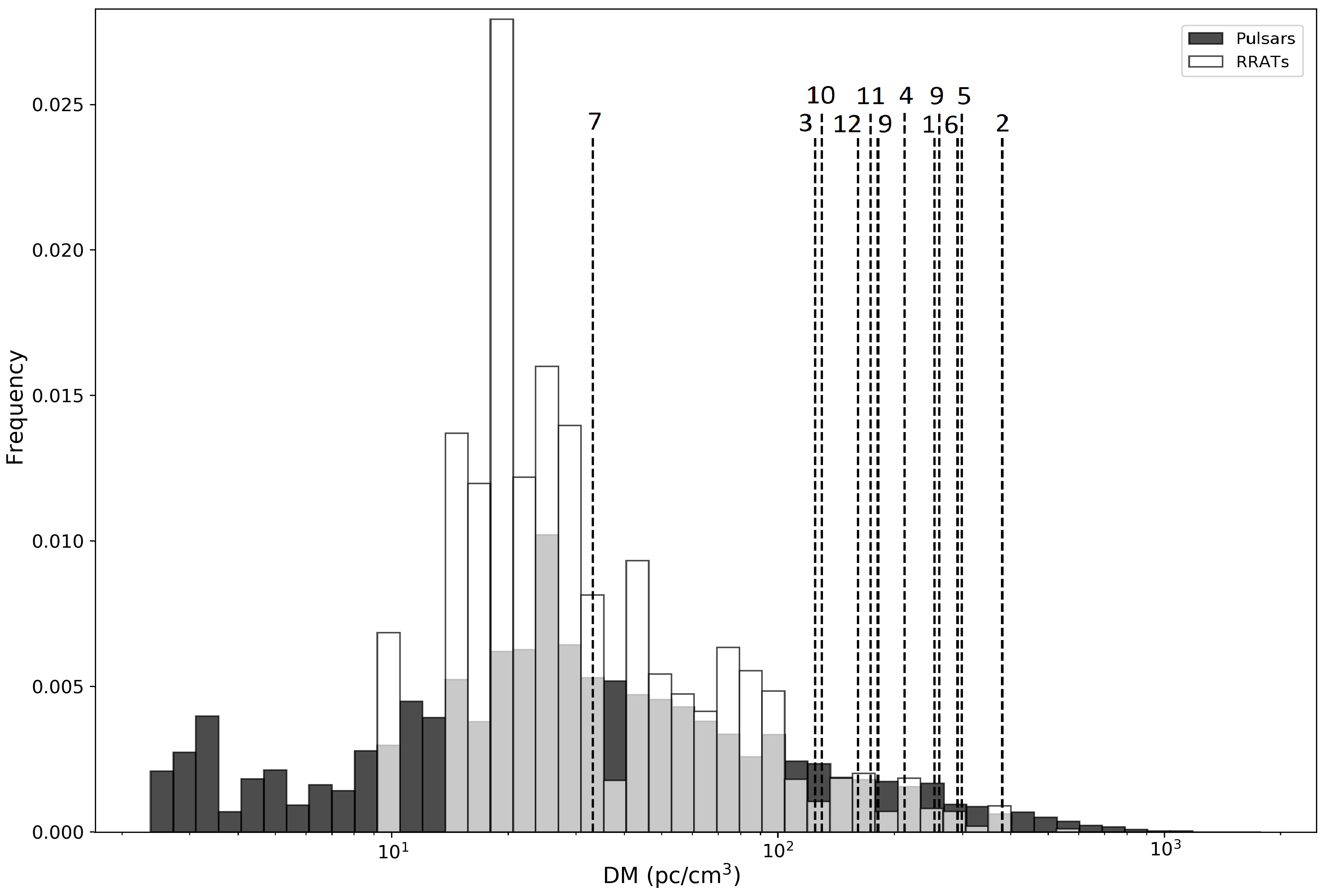}
    \caption{Distributions of DMs for known pulsars and RRATs compared to the MeerTRAP sources. The MeerTRAP sources are marked by the numbers corresponding to MTP0001--MTP0012, whose DMs are indicated by the dashed vertical lines. Note that the histogram bins are logarithmically sized, and that the histograms are normalised so that the area underneath each of them integrates to unity.}
    \label{fig:DMdistr}
\end{figure*}

\begin{figure}
    \centering
    \includegraphics[width=0.48\textwidth]{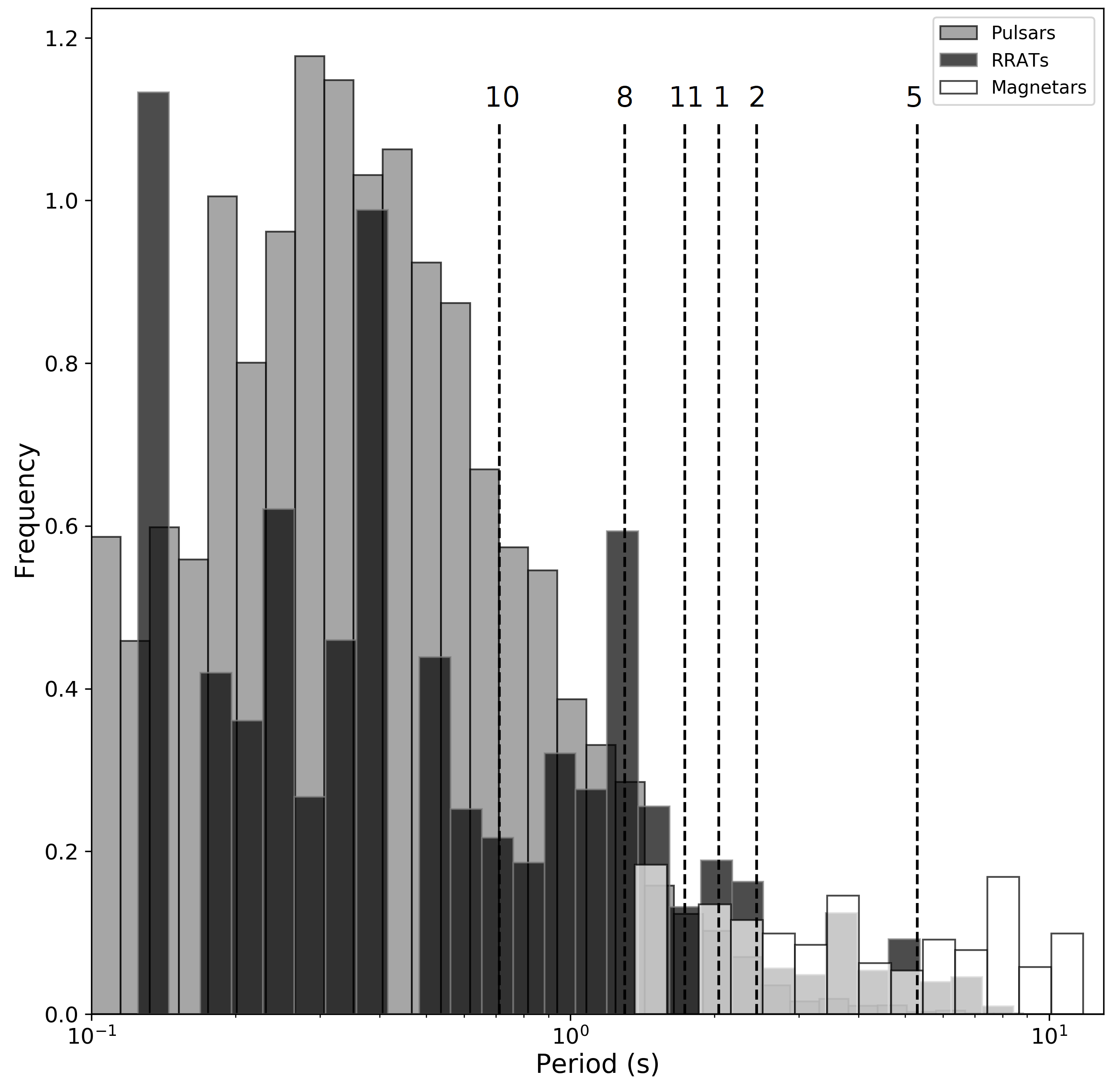}
    \caption{Distributions of pulse periods for known pulsars, RRATs, and magnetars, with the periods of the five MeerTRAP sources with measured periods---viz.~MTP0001/PSR~J1843$-$0757, MTP0002/PSR~J1152$-$6056, MTP0005/PSR~J1840$-$0840, MTP0008/PSR~J1901+0254, MTP0010/PSR~J0723$-$2050 and MTP0011/PSR~J0943$-$5305---shown by the dashed vertical lines. Millisecond pulsars (P < 0.1~s) have been excluded. Note that the histogram bins are logarithmically sized, and that the histograms are normalised so that the area underneath each of them integrates to unity.}
    \label{fig:Perdistr}
\end{figure}

\begin{table*}
\caption{Detection and timing parameters of the twelve new Galactic fast transient sources found by MeerTRAP. The observing times are in minutes. Note that the observing time for MeerKAT includes pointings for which the source coordinates were within the CB tiling region, while the Lovell and Parkes telescope observations were all targeted at the source. The period, period derivative, and epoch listed for PSR~J1840$-$0840 are taken from \citet[][]{2006lorimer}, and those of PSR~J1901+0254 are from \citet{2004hobbs}.}\label{table:refined_parameters}
\begin{tabular}{lrrrrrrlll}
\toprule
 & \multicolumn{2}{c}{MeerKAT} & \multicolumn{2}{c}{Lovell} & \multicolumn{2}{c}{Parkes} \\
\cmidrule(lr){2-3}\cmidrule(lr){4-5}\cmidrule(lr){6-7}
Source name & T$_{\mathrm{Obs}}$ & Pulses & T$_{\mathrm{Obs}}$ & Pulses & T$_{\mathrm{Obs}}$ & Pulses & P~(s) & $\Dot{\mathrm{P}}$ ($10^{-15}$) & Epoch\\
\midrule
PSR~J1843$-$0757 & 145 & 65 & 540 & 51 & 60 & 101 & 2.03194008516(9) & 4.13(3)  & 58743.79446\\
PSR~J1152$-$6056  & 114 & 9 & 0 & 0 & 120 & 0 & 2.449000(6) & & 58925.72910\\
MTP0003 & 9 & 1 & 46 & 0 & 30 & 0 & & \\
MTP0004 & 44 & 1 & 0 & 0 & 30 & 0 & & \\
PSR~J1840$-$0840 & 146 & 3 & 0 & 0 & 20 & 0 &  5.3093766847(20) & 23.7(12) & 53278.00000 \\
MTP0006 & 23 & 1 & 112  & 0 & 25 & 0 & & & \\
PSR~J0930$-$1854 & 250 & 2 & 126 & 0 & 55 & 0 & & \\
PSR~J1901+0254 & 215 & 107 & 11 & 0 & 30 & 0 & 1.2996934495(3) & 0.46(11)  & 52626.00000\\
MTP0009 & 97 & 1 & 89 & 0 & 90 & 0 & & & \\
PSR~J0723$-$2050 & 55 & 7 & 0 & 0 & 65 & 0 & $\sim$0.71154 & & 59165.22799\\
PSR~J0943$-$5305 & 98 & 7 & 0 & 0 & 85 & 0 & $\sim$1.734 & & 58925.72795 \\
MTP0012 & 26 & 1 & 0 & 0 & 0 & 0 & & & \\     
\bottomrule
\end{tabular}
\end{table*}

\begin{table}
\centering
\caption{Localisation coordinates of MeerTRAP Galactic sources, with 1-$\sigma$ errors. The positions for the sources marked by asterisks are from previous publications on the source, while the others are new localisations based on the single pulses detected by MeerTRAP. Note that for the latter localisations the 1-$\sigma$errors are statistical only (i.e.~arising from the model fitting of the S/N in different beams), and there may be other, systematic errors still unaccounted for.}\label{table:localisations}
\begin{tabular}{lrr}
\toprule
Source name & RA (hms) & Dec (dms)\\
\midrule
PSR~J1843$-$0757 & 18:43:33.06(2) & $-$07:57:33(2)\\
PSR~J1152$-$6056 & 11:52:37.44(13) & $-$60:56:22(1)\\
PSR~J1840$-$0840* & 18:40:51.9(4) & $-$08:40:29(15)\\
PSR~J0930$-$1854 & 09:30:35.27(2) & $-$18:54:31.1(8)\\
PSR~J1901+0254* & 19:01:15.67(7) & 02:54:41(5)\\
PSR~J0723$-$2050 & 07:23:04.8(18) & $-$20:49:52(7)\\
\bottomrule
\end{tabular}
\end{table}

\section{Discussion and conclusions}
\label{sec:conclusions}

We have described the first twelve Galactic sources that the MeerTRAP project has discovered operating our real-time, commensal search for fast radio transients to date. We have also outlined the follow-up observations and analysis performed on the discovered sources.

Figure~\ref{fig:DMdistr} shows the DMs of the twelve sources compared to those of known pulsars (as recorded in the ATNF pulsar catalog) and RRATs (from the RRATalog). It is apparent that the sources discovered by MeerTRAP tend to have higher DMs than the mean value for either pulsars or RRATs. All of the sources besides PSR~J0930$-$1854 are consistent with the higher end of RRAT DMs, which is to be expected considering the weighting of MeerTRAP observing time thus far towards the Galactic plane, where the column density of free electrons that disperse radio signals is highest\footnote{Although it is notable that the known RRAT population may itself be biased towards higher DM values, given the preponderance of transient searches that have been carried out close to the Galactic plane compared to those far from it.}. Furthermore, given MeerKAT's high sensitivity, MeerTRAP's single pulse detection has a lower flux density limit than preceding surveys, so that more distant sources are detectable.

Only one of the twelve sources has been detected enough times to establish a coherent timing solution, namely PSR~J1843$-$0757. For PSR~J1152$-$6056, PSR~J0723$-$2050, and PSR~J0943$-$5305 approximate pulse periods could be determined from a small number of pulses seen with MeerKAT. Figure~\ref{fig:Perdistr} compares the pulse periods of these three sources, as well as those of PSR~J1840$-$0840 and PSR~J1901+0254, which have known timing solutions from previous detections, to those of known pulsars (excluding millisecond pulsars), RRATs, and magnetars (from the McGill online magnetar catalog). These sources' periods would be typical for RRATs, towards the tail end of the distribution for canonical pulsars, and short for magnetars. Table~\ref{table:refined_parameters} contains a summary of all source detections as well as the timing parameters derived from them.

The first MeerTRAP discovery, PSR~J1843$-$0757, is the only one that we have found to be regularly repeating and observable. In all we have confirmed 218 pulses from the source during observations amounting to over 13 hours using the MeerKAT, Lovell, and Parkes telescopes. We have shown that the wait times between subsequent pulses from PSR~J1843$-$0757 do not follow an exponential distribution, and therefore are consistent with significant clustering in time.

We also obtained a coherent timing solution for PSR~J1843$-$0757, calculating a period of 2.03194008516(9)~s and period derivative of 4.13(3)$\times10^{-15}$, which has proven stable across 22 months of observations. Additionally, from the pulse timing we derived a precise localisation which agrees within 1-$\sigma$ errors with that found using \texttt{SeeKAT}. This proves the efficacy of the localisation method we applied to find best-fit coordinates at arcsecond-level precision for PSR~J1152$-$6056, PSR~J0930$-$1854, and PSR~J0723$-$2050. The best-fit coordinates derived with this method are listed in Table~\ref{table:localisations}.

Of the twelve sources presented in this paper, six have been once-off occurrences, and another two have never been re-detected subsequent to their discoveries. This may be because the sources are infrequently active, but it is also possible that these discoveries represent pulses on the bright end of the sources' pulse flux distributions that occur more rarely, and so are less likely to be seen in follow up observations, especially with less sensitive telescopes.

An important factor to bear in mind regarding the MeerKAT observations on a particular MeerTRAP source is that little of that time would have been spent pointed directly towards the source coordinates, but rather somewhere else such that the source was located in a portion of the MeerKAT primary beam with reduced sensitivity. As an illustration of how this affects source detection rates, the detection rate of PSR~J1843$-$0757 with the Parkes telescope was comparable to that when MeerKAT was trained directly on the source coordinates (101 pulses in 60 minutes compared to 65 pulses in 40 minutes\footnote{In table \ref{table:refined_parameters} the quoted MeerKAT observing time also includes observations when the source coordinates were within the CB tiling region but not at boresight.}). However, no PSR~J1843$-$0757 pulses were detected during the further $\sim$100 minutes when the source was within the CB tiling region but not at the boresight. Here it is important to bear in mind that the standard CB overlap level is 25 per cent; hence MeerTRAP's sensitivity for a given observation could vary by up to a factor of four depending on the source's position within the nearest CB.

Furthermore, we only had a timing solution for PSR~J1843$-$0757 and periods determined over short data spans for PSR~J1152$-$6056, PSR~J0723$-$2050, and PSR~J0943$-$5305. Those sources' data could be folded at the known period and the additive result inspected for evidence of repeating pulses. For the other eight sources that we had commissioned Parkes or Lovell observations of, the data had to be searched for individual single pulses, reducing the chances of detection. Additionally, both the Lovell and Parkes datasets were heavily affected by RFI, which had to be excised using algorithms that could have degraded the S/N values of pulses present in the data. Lastly, PSR J1840$-$0840 was detected in the MeerKAT IB alone, and follow up observations were pointed at the IB boresight coordinates, despite the true source position later being determined as $\sim$50\arcmin~from this position. For these reasons we caution against over-interpreting the rate of positive detections.

The case of MTP0008/PSR~J1901+0254, which was unexpectedly detected by MeerTRAP despite the fact that previous average flux  density measurements would put the source significantly below the MeerTRAP detection threshold, is intriguing and encourages follow up. Further, long term monitoring is required to establish both possible flux density evolution over time and the nature of the large pulse-to-pulse variability. 

Finally, the initial confusion of both MTP0005/PSR~J1840$-$0840 and MTP0008/PSR~J1901+0254 for new discoveries also serves to emphasise the importance of obtaining accurate DM measurements for future single pulse searches. Since the number of known transients is likely to increase markedly once very sensitive telescopes like the SKA commence observations, precise DM figures will be of great value for distinguishing discoveries from known sources. It is also apparent that a future version of a pulsar catalogue will need to contain more information on the sources, such as flux variability and emission properties such as drifting subpulses to ease the process of determining whether sources are new. As demonstrated by \citet[][]{2017cooper}\footnote{PhD thesis available online at \url{https://ethos.bl.uk/OrderDetails.do?uin=uk.bl.ethos.756836}}, we are already in the regime where the population of pulsars is large enough that there is a reasonable probability of pulse periods being the same for a few pairs of pulsars to 4 significant figures.

\section*{Acknowledgements}
The authors would like to thank the reviewer for their useful suggestions. We also heartily thank the MeerKAT LSP teams for allowing us to observe commensally. We are indebted to Sarah Buchner, Maciej Serylak, and Dave Horn at the South African Radio Astronomy Observatory for their support of the MeerTRAP project. We also thank the staff of the Lovell and Parkes telescopes for their help in carrying out the follow up observations. This project has received funding from the European Research Council (ERC) under the European Union’s Horizon 2020 research and innovation programme (grant agreement No. 694745). The MeerKAT telescope is operated by the South African Radio Astronomy Observatory, which is a facility of the National Research Foundation, an agency of the Department of Science and Innovation. Pulsar research at Jodrell Bank Centre for Astrophysics and Jodrell Bank Observatory is supported by a consolidated grant
from the UK Science and Technology Facilities Council (STFC). The Parkes radio telescope is part of the Australia Telescope National Facility which is funded by the Australian Government for operation as a National Facility managed by CSIRO. We acknowledge the Wiradjuri people as the traditional owners of the Parkes Observatory site.


\section*{Data Availability}
The data will be made available to others upon reasonable request
to the authors.



\bibliographystyle{mnras}
\bibliography{Main} 







\bsp	
\label{lastpage}
\end{document}